\documentclass[aps,prl,preprint,amsmath,amssymb,superscriptaddress,nofootinbib]{revtex4-2}

\usepackage{braket,gensymb,tikz}
\usepackage{xcolor}
\usepackage{soul}

\newcommand{\comment}[1]{\iffalse {#1} \fi}

\begin{document}
\title{Dipolar spin-exchange and entanglement between molecules in an optical tweezer array}

\author{Yicheng Bao}
\author{Scarlett S. Yu}
\author{Lo\"ic Anderegg}
\affiliation{Department of Physics, Harvard University, Cambridge, MA 02138, USA}
\affiliation{Harvard-MIT Center for Ultracold Atoms, Cambridge, MA 02138, USA}

\author{Eunmi Chae}
\affiliation{Department of Physics, Korea University, Seongbuk-gu, Seoul 02841, South Korea}

\author{Wolfgang Ketterle}
\affiliation{Harvard-MIT Center for Ultracold Atoms, Cambridge, MA 02138, USA}
\affiliation{Department of Physics, Massachusetts Institute of Technology, Cambridge, MA 02139, USA }

\author{Kang-Kuen Ni} 
\affiliation{Department of Physics, Harvard University, Cambridge, MA 02138, USA}
\affiliation{Harvard-MIT Center for Ultracold Atoms, Cambridge, MA 02138, USA}
\affiliation{Department of Chemistry and Chemical Biology, Harvard University, Cambridge, MA 02138, USA}

\author{John M. Doyle} 
\affiliation{Department of Physics, Harvard University, Cambridge, MA 02138, USA}
\affiliation{Harvard-MIT Center for Ultracold Atoms, Cambridge, MA 02138, USA}

\date{\today}

\begin{abstract}
Due to their intrinsic electric dipole moments and rich internal structure, ultracold polar molecules are promising candidate qubits for quantum computing and for a wide range of quantum simulations. Their long-lived molecular rotational states form robust qubits while the long-range dipolar interaction between molecules provides quantum entanglement. Using a molecular optical tweezer array, single molecules can be moved and separately addressed for qubit operations using optical and microwave fields, creating a scalable quantum platform. Here, we demonstrate long-range dipolar spin-exchange interactions in pairs of CaF molecules trapped in an optical tweezer array. We control the anisotropic interaction and realize the spin-$\frac{1}{2}$ quantum XY model by encoding an effective spin-$\frac{1}{2}$ system into the rotational states of the molecules. We demonstrate a two-qubit (two-molecule) gate to generate entanglement deterministically, an essential resource for all quantum information applications. Employing interleaved tweezer arrays, we demonstrate high fidelity single site molecular addressability. 
\end{abstract}

\maketitle
\section{Introduction}

The long-range and spatially anisotropic electric dipole-dipole interaction between polar molecules make them a powerful platform for realizing quantum simulations of strongly interacting many-body dynamics~\cite{micheli2006toolbox,gadway2016strongly, baranov2008theoretical, wall2015magnetism} and for scalable quantum computing~\cite{demille2002quantum,yelin2006dipolarQC, karra2016paramagnetic,sawant2020qudits,ni2018dipolar}. These scientific goals require ultracold molecules with full quantum control of the dipole-dipole interaction to construct practical quantum information systems~\cite{carr2009coldmolecules}. Ultracold molecules can be produced either via direct laser cooling~\cite{barry2014magneto,anderegg2017radio,truppe2017molecules,collopy20183d,vilas2022CaOHMOT} or assembly of individual ultracold atoms~\cite{ni2008high,molony2014creation,takekoshi2014ultracold,park2015ultracold,guo2016creation,rvachov2017long,liu2018building}. Recent experimental progress has established the capability of preparing and manipulating ultracold molecules with high fidelity both in bulk gas or 3D lattices~\cite{yan2013observation,seesselberg2018extending,williams2018magnetic,li2022tunable} and individual tweezer or lattice sites~\cite{burchesky2021rotational,cairncross2021assembly,christakis2022probing}. Rearrangeable tweezer arrays offer an attractive platform due to their scalability and potential for single site addressability~\cite{endres2016atom,barredo2016atom,bernien2017probing,liu2018building, anderegg2019optical,zhang2022optical}. Molecules in optical tweezers have previously demonstrated long rotational and hyperfine state coherence times, significantly longer than the predicted molecule-molecule dipolar gate times~\cite{park2017second,seesselberg2018extending, burchesky2021rotational,gregory2021robust}.

A key milestone towards using ultracold polar molecules for quantum simulations and multi-particle quantum gates is the generation of coherent dipole-dipole couplings between molecules~\cite{gorshkov2011tunable,ni2018dipolar}, which has been shown in sparesely filled 3D lattice~\cite{yan2013observation,christakis2022probing} and in 2D layers~\cite{tobias2022reactions}. In this work, resonant electric dipole-dipole coupling is used for coherent exchange of rotational angular momentum between pairs of individually trapped laser-cooled CaF molecules in optical tweezers, parallel to concurrent work~\cite{holland2022demand}. With the ability to tune the angle of the molecular quantization axis in the lab frame, our system realizes both ferromagnetic and anti-ferromagnetic couplings in a quantum XY spin-exchange model, by effectively encoding a spin-$\frac{1}{2}$ system into molecular rotational states. Using this spin-exchange Hamilitonian, we perform iSWAP two-qubit gate operations. The iSWAP gate, when sequentially combined with single-qubit operations, can generate bipartite entanglement deterministically and form a universal set of quantum gates, which is an essential resource for all quantum information applications~\cite{turchette1998deterministic,ni2018dipolar}. We find that the fidelity of generated Bell states is limited by thermal motion of molecules within the tweezer traps. Finally, we demonstrate the use of an interleaved dual tweezer array system, which allows for robust single site addressability and fast gate operations between molecules. This, combined with expected future molecular cooling~\cite{caldwell2020sideband}, has the potential to greatly increase the fidelity of two-qubit operations in this system.

\section{Initial State Preparation of CaF Molecules in a 1D Optical Tweezer Array}
The experiment starts by loading CaF molecules from a cryogenic buffer gas beam source~\cite{hutzler2012buffer} into a radio-frequency magneto-optical trap~\cite{anderegg2017radio}. Next, molecules are loaded into a one-dimensional (1D) optical lattice using $\Lambda$-enhanced gray molasses cooling on the electronic $X-A$ transition~\cite{cheuk2018lambda, anderegg2018laser}, and optically transported into a glass cell using a focus-tunable moving lattice, characterized in our previous work~\cite{bao2022fast}.

\begin{figure}[!htbp]
\includegraphics[width=\textwidth]{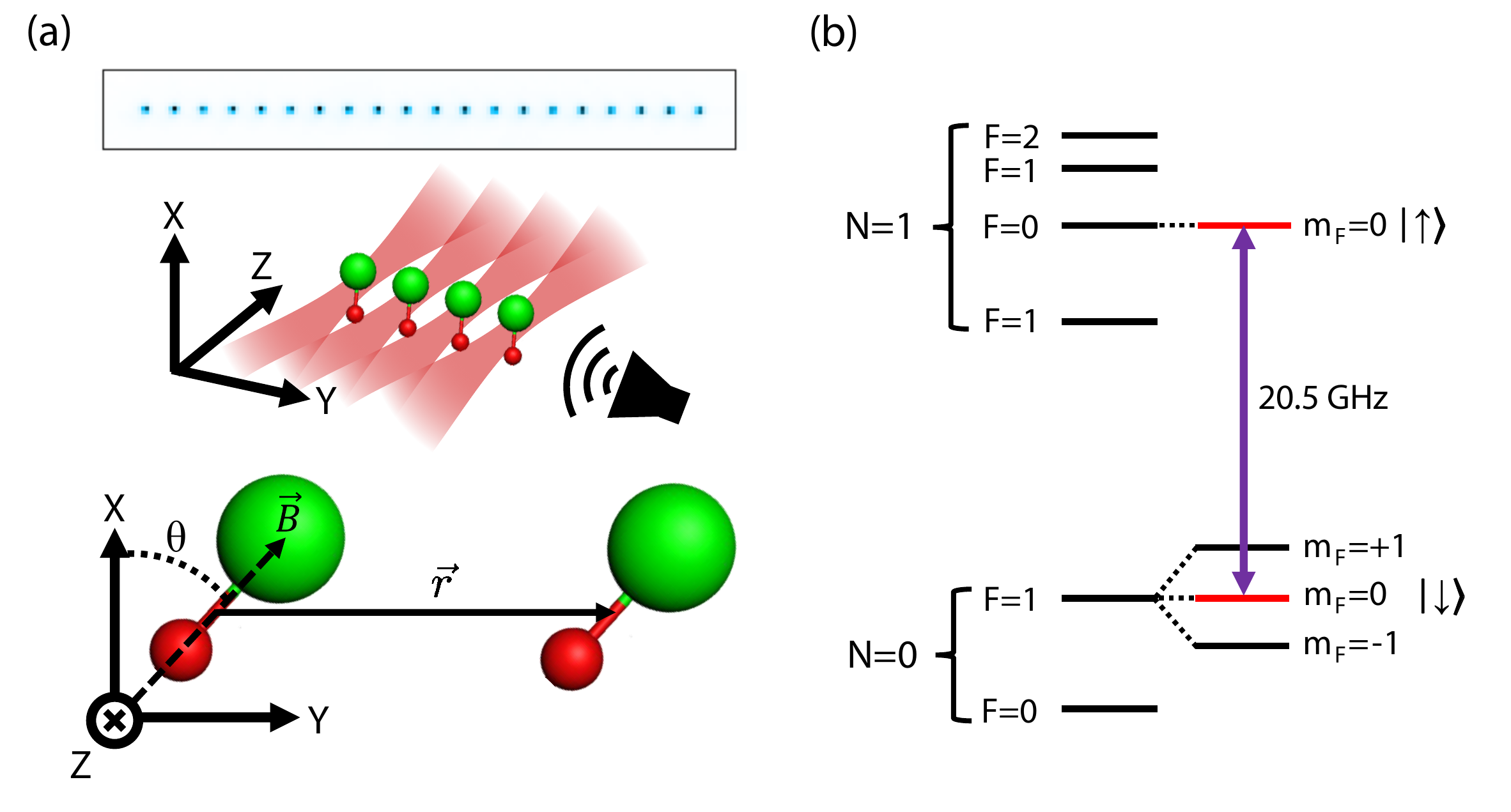}
\caption{(a) (Top) The image of averaged flourescence from a 20-site optical tweezer array of CaF molecules. (Middle) Cartoon of optical tweezers and the coordinate system used in this work. For visual clarity, only four sites of the realized twenty-site array are depicted. (Bottom) An illustration showing the relative angle between the applied bias magnetic field and the tweezer array arrangement. The tweezer light polarization is along Z (perpendicular to the bias magnetic field). The $|\vec{r}|$ denotes the instantaneous spacing between molecules. (b) Relevant rotational ($N$) and hyperfine ($F$) states in the ground electronic state of CaF.}
\label{fig:1}
\end{figure}

A microscope objective with 0.6 numerical aperture is used to project a 1D optical tweezer array in the glass cell, as well as collect fluorescence from the molecules during $\Lambda$-imaging~\cite{cheuk2018lambda}. The array is formed by passing single-frequency $776\,\text{nm}$ laser light through a shear-mode acousto-optic deflector (AOD) driven with a multi-tone waveform~\cite{anderegg2019optical}, which is generated by a high-speed arbitrary waveform generator (AWG). This allows for control of the position of individual optical tweezer sites as well as trap depth. We start with a twenty-site array formed by a single AOD (Fig.~\ref{fig:1}~(a)).

We load the molecules into the tweezers by overlapping the array with a transported molecular cloud that is held in an optical dipole trap in the presence of the $\Lambda$-cooling light. We observe an average probability of $35\%$ for loading a single site with a single molecule. After loading the array, we apply a $\Lambda$-imaging pulse and collect fluorescence onto a camera to identify the tweezer sites that are loaded with single molecules in the ground electronic and vibrational state (or empty) \cite{cheuk2018lambda,anderegg2019optical}. Since the $\Lambda$-cooling technique relies on closed photon cycling between the $\ket{X,N=1}$ and $\ket{A,J=1/2}$ manifolds, only molecules in the $\ket{X,N=1}$ rotational manifold can be loaded into the tweezers and detected in this phase of the experiment.

Imaging distributes the population of molecules over all twelve hyperfine states. In order to prepare the molecules in a single quantum state, we use a combination of optical pumping and microwave transfer. Using short $X-A$ laser pulses resonant with all hyperfine levels in $\ket{X,v=0,N=1}$ manifold {\emph{except}} the $\ket{X,N=1,F=0,m_F=0}$ state, we pump most of the molecular population into the latter state. The trap depth of the tweezers is then linearly ramped down in $5\,\text{ms}$ and a bias magnetic field of $\approx 3.2\,\text{G}$ is applied. A subsequent microwave $\pi$-pulse transfers the population from $\ket{X, N=1,F=0,m_F=0}$ to the $\ket{X, N=0,F=1,m_F=0}$ state (Fig.~\ref{fig:1}~(b)). Finally, an $X-A$ laser pulse containing frequencies to drive all the hyperfine components is applied to remove any  molecular population left over in the $\ket{X,N=1}$ manifold. In the end, all the remaining molecules in the array are in the $\ket{X, N=0,F=1,m_F=0}$ state, initializing the qubits, and effectively encoding a spin-$\frac{1}{2}$ model in the subspace spanned by $\ket{X, N=1,F=0,m_F=0} \equiv \ket{\uparrow}$ and $\ket{X, N=0,F=1,m_F=0} \equiv \ket{\downarrow}$ states (Fig.~\ref{fig:1}~(b)). The total state preparation and detection efficiency is $60(2)\%$, limited by residual population in other hyperfine states within $\ket{X, N=1}$ and imperfect imaging fidelity.

\section{Single molecule rotational coherence time}
To observe high-fidelity dipolar spin-exchange interactions, a long rotational coherence time comparable to the time scale of the dipolar interaction is required. With a sample of molecules at finite temperature in optical tweezers, it is important to control the differential AC stark shift broadening due to molecule's thermal motion in the tweezer trap \cite{neyenhuis2012anisotropic,seesselberg2018extending,chae2021entanglement,lin2021anisotropic}. To suppress this broadening, the tweezer laser is linearly polarized at a ``magic" angle relative to the quantization axis defined by the applied bias magnetic field, as detailed previously \cite{burchesky2021rotational} (Fig.~\ref{fig:1}~(a)). To maximize the single qubit coherence time, it is beneficial to adiabatically lower the trap depth as much as possible without unduly spilling molecules from the trap. However, this is not optimal for maximizing the number of observed dipolar spin-exchange oscillation cycles. At finite temperature, the instantaneous dipolar interaction strength fluctuates due to thermal motion, which becomes more prominent in a shallow trap. This is the main mechanism of dephasing. To mitigate this issue, we confine the molecules more tightly by operating the tweezer at a higher trap light intensity for which the magic angle is close to $90\degree$. The work described below is performed under these conditions.

We measure the rotataional qubit coherence time under conditions of three different microwave pulse sequences used to drive transitions between $\ket{\uparrow}$ and $\ket{\downarrow}$ and  observing the population in $\ket{\uparrow}$. With a Ramsey sequence, we observe a single qubit coherence time of $\tau_C = 3.6(6)\,\text{ms}$. By adding a single $\pi$ spin-echo pulse, $\tau_C$ is extended to $33(5)\,\text{ms}$. Active magnetic field cancellation is employed to remove long-term drift on the order of $10\,\text{mG}$, but the system used does not remove magnetic field fluctuation within a power line cycle time of less than  $1/60\,\text{Hz}$. This limits the effective interval between spin-echo pulses to be an integer multiple of $60\,\text{Hz}$, making it difficult to measure fast dipolar oscillations with only a single $\pi$-pulse. Instead, we use dynamical decoupling schemes to preserve the qubit coherence. This technique is employed in a variety of quantum information systems \cite{gullion1990new,du2009preserving}, including molecular systems \cite{yan2013observation,li2022tunable,holland2022demand}. We choose the XY8 dynamical decoupling sequence (Fig.~\ref{fig:2}~(a)), with a cycle length of $1.6\,\text{ms}$\footnote{The XY8 sequence effectively acts as a band-pass filter with, in our case, its pass-band centered at $10\,\text{kHz}$ \cite{choi2020robust}, which is higher than the dominant magnetic field noise spectrum range in our system.}, much shorter than  $1/60\,\text{Hz}$, achieving a coherence time of $\tau_C = 630(90)\,\text{ms}$. This dynamical decoupling is used for all of our measurements except where noted\comment{(see supplemental material)}. Fig. 2(b) shows the measured contrast versus time for single particle oscillations between $\ket{\uparrow}$ and $\ket{\downarrow}$, from which $\tau_C$ is determined.

\section{Coherent dipolar spin-exchange interaction}
The dipolar spin-exchange interaction Hamiltonian~\cite{gorshkov2011tunable}~is
\begin{equation}
H_{dip}=\frac{J}{2}(\hat{S}^+_1\hat{S}^-_2+\hat{S}^-_1\hat{S}^+_2)=J(\hat{S}^x_1\hat{S}^x_2+\hat{S}^y_1\hat{S}^y_2)\\
\label{eq:dipolar_hamiltonian}
\end{equation}
where $\hat{S}^+_i$ ($\hat{S}^-_i$, $\hat{S}^x_i$, $\hat{S}^y_i$) is the spin-$\frac{1}{2}$ raising (lowering, Pauli-X, Pauli-Y) operator for molecule number $i$ in a tweezer pair. $J$ is the dipolar interaction strength that can be further expressed as
\begin{equation}
J=\frac{d^2}{4 \pi \epsilon_0 r^3}(1-3 \cos^2\theta)
\label{eq:dipolar_J}
\end{equation}
where $d$ is the transition dipole moment between the $\ket{\uparrow}$ and $\ket{\downarrow}$ state ($d\approx1\,\text{Debye}$), $\epsilon_0$ is the vacuum permittivity, $r$ is the inter-molecular spacing and $\theta$ is the angle between the quantization axis and the inter-molecular axis direction.

We set the bias magnetic field perpendicular to both the $k$-vector of the tweezer light and the direction of 1D tweezer array. The polarization of the linearly polarized tweezer light is rotated to the tweezer array direction (i.e. a magic angle close to $90\degree$). This configuration provides the largest ``magic trap depth" (the tightest confinement) at a given magnetic field.

State preparation is first done with pairs of tweezers spaced at $|\vec{R}| \sim 5\,\mu\text{m}$, with each site loaded with a single molecule prepared in the $\ket{\downarrow}$ state (or empty). At this separation, the dipolar interaction strength $J$ is negligible ($\frac{J}{h}<3\,\text{Hz}$). In a time period $ \sim 1\,\text{ms}$ we then move the even numbered sites towards the odd numbered sites, by sweeping the AOD frequency tones of the even sites, which reduces the separation and has the effect of increasing the dipolar interaction strength. A $\frac{\pi}{2}$-pulse is then applied to prepare both molecules in the superposition state $(\ket{\uparrow}+\ket{\downarrow})/\sqrt{2}$. Then, we apply XY8 dynamical decoupling sequence of microwave pulses. Under the time evolution of dipolar spin-exchange Hamiltonian ($H_{dip}$), a relative phase accumulates between $(\ket{\downarrow\uparrow}+\ket{\uparrow\downarrow})/\sqrt{2}$ and 
$(\ket{\uparrow\uparrow}+\ket{\downarrow\downarrow})/\sqrt{2}$. (Since the XY8 sequence only contains $\pi$-pulses, it will not affect the phase accumulation during the evolution under $H_{dip}$.) After a wait time, we apply another $\frac{\pi}{2}$-pulse and then move the molecules apart. To read out the final qubit state, a second $\Lambda$-imaging pulse projects the system to $\ket{\uparrow\uparrow}$. We post-select on the data, pairs where both sites are initially loaded with single molecules in the $\ket{\downarrow}$ state and our measurement yields the probability $P_{\uparrow\uparrow}$ of detecting both molecules being in the $\ket{\uparrow}$ state. To summarize, starting from the initial state $\ket{\downarrow\downarrow}$, a microwave pulse sequence creates a final state that evolves in time as $\ket{\psi(t)}=\frac{1}{2}((1-e^{-i\frac{Jt}{2\hbar}})\ket{\downarrow\downarrow}-(1+e^{-i\frac{Jt}{2\hbar}})\ket{\uparrow\uparrow})$, resulting in the probability $P_{\uparrow\uparrow}=\cos^2{\frac{Jt}{4\hbar}}=\frac{1}{2}(1+\cos{\frac{Jt}{2\hbar}})$, which oscillates at an angular frequency of $\omega_{J} = \frac{J}{2\hbar}$.

\begin{figure}[!htbp]
\includegraphics[width=\textwidth]{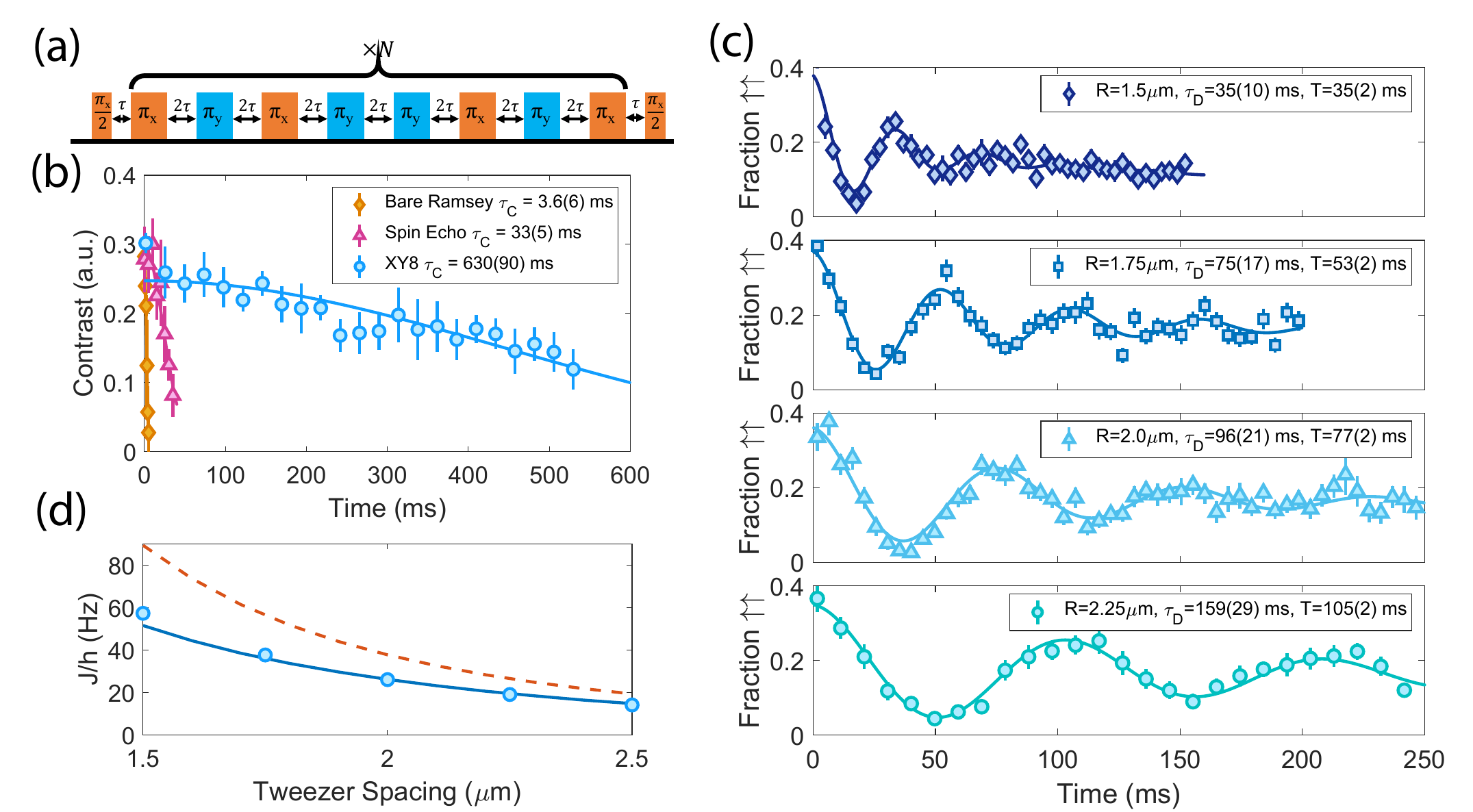}
\caption{(a) Ramsey sequence with XY8 dynamical decoupling used in this work. $\frac{\pi_x}{2}$ denotes a $\frac{\pi}{2}$-pulse, which rotates the quantum state around x-axis of the Bloch sphere by $90\degree$. Similarly, $\pi_x$ ($\pi_y$) denotes a $180\degree$ rotation around x-axis (y-axis) on the Bloch sphere. ``$\times N$'' means the block of XY8 pulses is repeated multiple times during the evolution time. (b) Measured single molecule rotational coherence time between $\ket{\uparrow}$ (N=1)  and $\ket{\downarrow}$ (N=0) using Ramsey, single $\pi$ pulse spin-echo, and the XY8 dynamical decoupling sequences. The coherence time is fitted as a Gaussian $1/e$ decay time constant. (c) Dipolar spin-exchange oscillation at various tweezer spacing, with fitted decay time constant and oscillation period shown in the legends. (d) Dipolar spin-exchange interaction strength $J$ versus the tweezer spacing $|\vec{R}|$. Orange line (dashed) is the theoretical prediction of $J$ at zero temperature. Blue line (solid) is the simulated result of $J$ with the thermal motion of the molecules taken into account.}
\label{fig:2}
\end{figure}

At smaller tweezer spacings, we observe increased $\omega_{J}$ due to the stronger dipolar interaction between the two molecules (Fig.~\ref{fig:2}~(c)). By fitting the data to an exponentially decaying sinusodial model, we extract the dipolar oscillation cycle period $T$ and contrast decay time constant $\tau_D$. The dipolar spin-exchange strength $J$ at different tweezer spacings can be then calculated from $T$ (Fig.~\ref{fig:2}~(d)). We find that the measured $J$ is slightly smaller than the theoretical prediction and deviates more as the spacing decreases. This can be explained by the finite temperature of the molecules causing the effective inter-molecular spacing $\braket{|\vec{r}|}$ to be larger than the tweezer spacing $|\vec{R}|$. We Monte-Carlo simulate the behavior of the molecules in the tweezer, including the thermal motion of the molecules, and show in Fig.~\ref{fig:2}~(d) that the simulated results agree with the experimental data.

\section{Anisotropy of the Dipolar Interaction}
The general dipole-dipole Hamiltonian described in Eq.~\ref{eq:dipolar_hamiltonian} and Eq.~\ref{eq:dipolar_J} is inherently anisotropic due to the $\theta$-dependent term. We study the effect of the anisotropy of spin exchange experimentally by varying the angle $\theta$, and measuring the corresponding dipolar interaction strength. The quantization axis is set by the applied bias magnetic field and can be changed by tuning the current through two pairs of magnetic field coils while simultaneously rotating the tweezer light polarization to maintain the same magic angle.

\begin{figure}[!htbp]
\includegraphics[width=0.7\columnwidth]{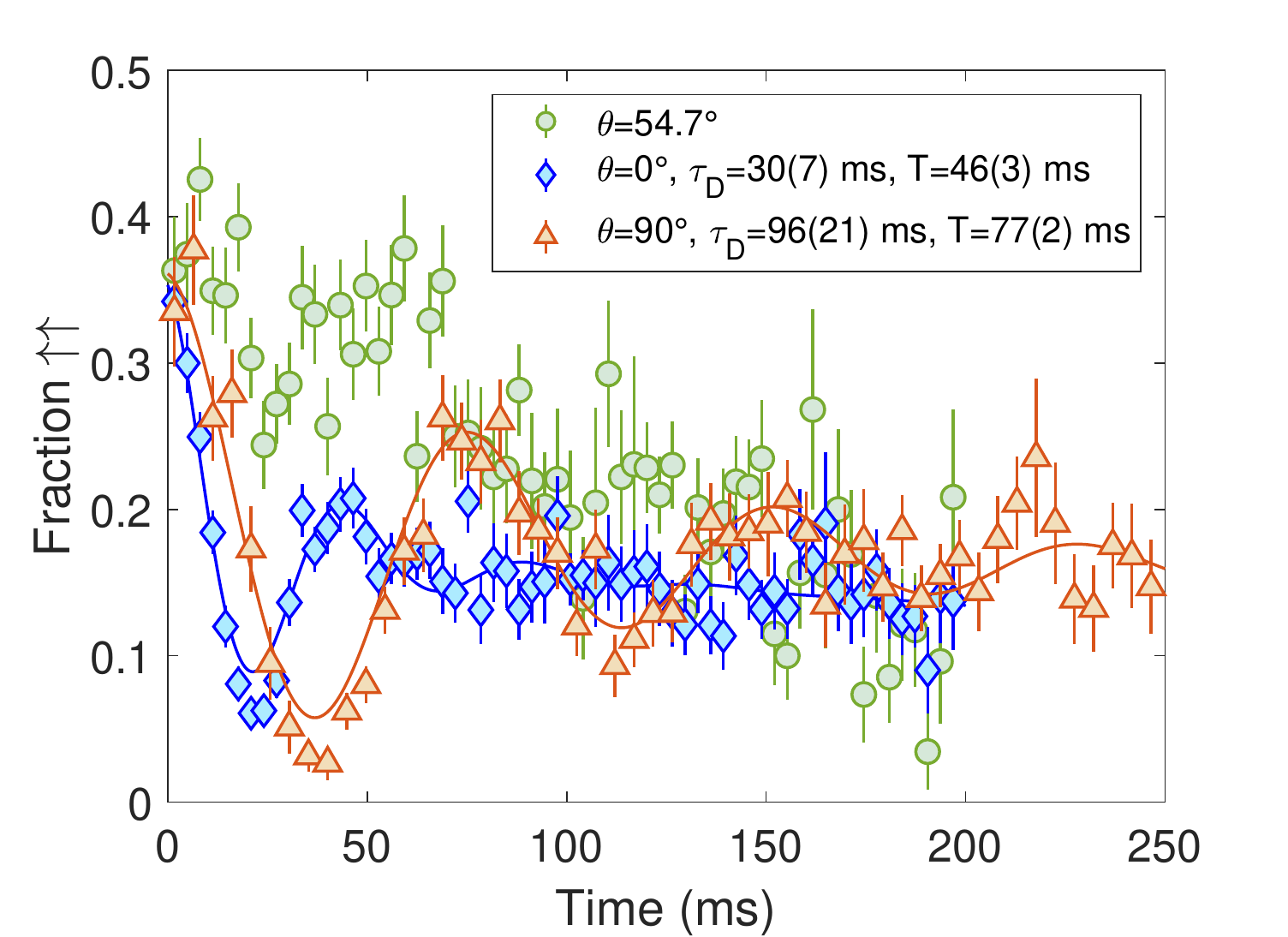}
\caption{Dipolar spin-exchange interaction at different angles $\theta=0\degree$, $54.7\degree$ and $90\degree$. The $\theta=0\degree$ and $90\degree$ data is fitted to an exponentially decaying sinusodial model with decay time constant $\tau_D$ and oscillation cycle time $T$, shown in solid lines. \comment{Simulated result are shown in dashed lines.}}
\label{fig:angular_dependence}
\end{figure}

In Fig.~\ref{fig:angular_dependence}, we show $P_{\uparrow\uparrow}$ at three characteristic angles $\theta=0\degree$, $54.7\degree$ and $90\degree$, all taken with a $|\vec{R}|=2\,\mu\text{m}$ tweezer spacing. We observe dipolar spin-exchange oscillations in both $\theta=0\degree$ and $\theta=90\degree$ configurations, with the oscillation at $\theta=0\degree$ being twice the frequency as that of $\theta=90\degree$. No clear oscillation is observed for $\theta=54.7\degree$ configuration, as expected because the dipolar interaction averages to zero at this angle. These results agree with the absolute value of the magnitude of the anisotropic term in each configuration. We also observe a larger number of oscillation cycles at $\theta=90\degree$ than $\theta=0\degree$. Our Monte-Carlo simulation indicates that thermal motion dephasing is the dominant cause. At $\theta=0\degree$, the larger motional wavefunction spread in the more weakly trapped axial direction of the optical tweezers results in a large fluctuation of the instantaneous value of $\theta$\comment{(see supplemental material)}. The $\theta=90\degree$ configuration is less affected due to the tight confinement of the optical tweezers in the radial direction.

\section{Fidelity of Created Bell States}

\begin{figure}[!htbp]
\includegraphics[width=0.7\columnwidth]{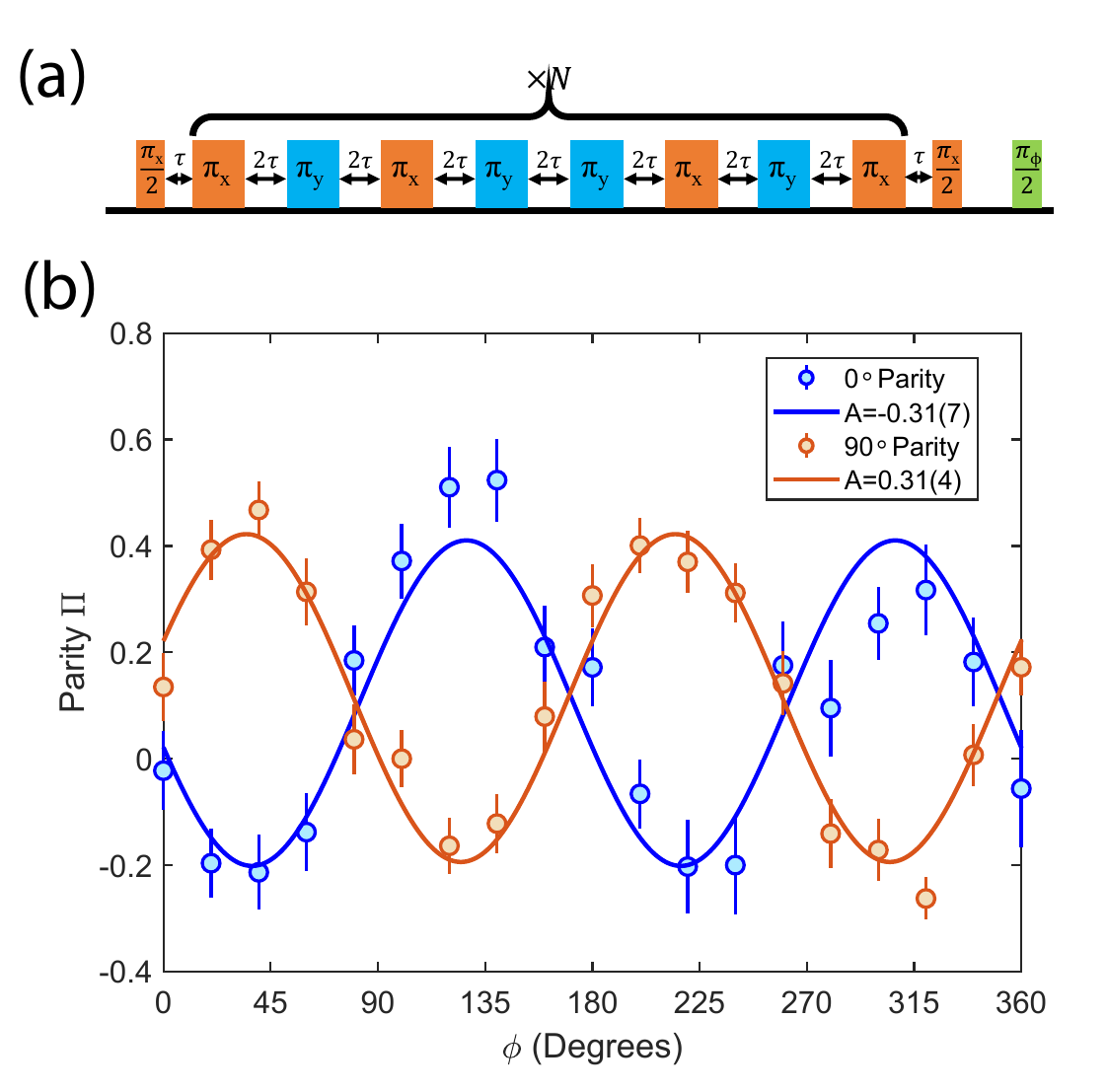}
\caption{{\textbf{Parity measurements showing creation of Bell state pairs}} (a) Microwave pulse sequence used in the parity oscillation measurement. $\frac{\pi_\phi}{2}$ denotes a $\frac{\pi}{2}$-pulse with a microwave phase shifted by $\phi$ relative to the first $\frac{\pi_x}{2}$ pulse in the sequence, effectively rotating the quantum state around an axis that is angled $\phi$ relative to the x-axis on the Bloch sphere. (b) Parity oscillation at $\theta=0\degree$ and $\theta=90\degree$. $A$ is the fitted parity oscillation amplitude.}
\label{fig:4}
\end{figure}

Dipolar spin-exchange can be used as a two-qubit gate to generate entanglement between molecules in neighboring tweezers~\cite{ni2018dipolar}. At a dipolar spin-exchange interaction time of~$t=~T/4=19.2\,\text{ms}$, where $T$ is the oscillation period of the spin-exchange oscillation, the system has evolved into a maximally entangled state known as a Bell state. To test the fidelity of the Bell state generated in our system, we apply a third $\frac{\pi}{2}$-pulse around a variable rotation axis on the Bloch sphere (angled $\phi$ relative to the x-axis on the equatorial plane) (Fig.~\ref{fig:4}~(a)). By varying $\phi$ and measuring the survival probability of all four possible final state outcomes, one can construct the parity quantity $\Pi$ \cite{turchette1998deterministic,sackett2000experimental}.
\begin{equation}
\Pi=\braket{\hat{\Pi}}=\braket{\hat{S}^z_1{\hat{S}^z_2}}=P_{\uparrow\uparrow}+P_{\downarrow\downarrow}-P_{\uparrow\downarrow}-P_{\downarrow\uparrow}
\end{equation}
Here $\braket{}$ denotes averaging over all occurrences post-selected pairs that are both loaded with a single molecule. $\hat{S}^z_i$ represents the Pauli-Z operator on the number $i$ molecule in a pair.

Starting from a Bell state, this sequence will result in a $4\pi$ oscillation in $\Pi$ as $\phi$ is varied from $0$ to $2\pi$ \cite{pezze2018quantum}. Displayed in Fig.~\ref{fig:4}~(b) is $\Pi$ for both $\theta=0\degree$ and $\theta=90\degree$ configurations. Extracting the contrast of the oscillation for the
$\theta=90\degree$ case, we measure the Bell state fidelity of $\mathcal{F}=0.31(4)$ and a state preparation and measurement (SPAM) corrected fidelity $\mathcal{F}_{SPAM}=0.86(13)$. The phase of the parity oscillation also reveals the sign of the anisotropic term $(1-3 \cos^2\theta)$. Our data shows that the $\theta=0\degree$ configuration leads to a negative $J$, corresponding to a ferromagnetic interaction, and the $\theta=90\degree$ configuration leads to a positive $J$, corresponding to an anti-ferromagnetic interaction.

\section{Towards arbitrary initial state preparation}

Motivated by the desire to perform robust single site addressing, instead of using a single AOD to generate all sites in the array, we switch to using one AOD to generate the odd numbered sites and another AOD to generate the even number sites. This allows for convenient independent trap depth control over each molecule in a pair, and uniformity across the array. Additionally, by offseting the frequency of the tweezer light of the even and odd sites, molecules can be moved in close proximity without experiencing the heating that can arise in a single AOD system~\cite{endres2016atom, zhang2022optical}.  For a given trap depth, the differential AC stark shift results in the molecules in even number sites being away from the resonance of the $ \ket{\uparrow} \rightarrow \ket{\downarrow}$ microwave transition, therefore allowing separate microwave addressing of the odd sites. By applying a microwave $\pi$-pulse when odd sites are detuned away, we can prepare an anti-ferromagnetic initial state $\ket{\uparrow\downarrow}$. Under $H_{dip}$, an initial state $\ket{\uparrow\downarrow}$ evolves as $\ket{\psi(t)}=\cos{\frac{Jt}{2\hbar}}\ket{\uparrow\downarrow}-i\sin{\frac{Jt}{2\hbar}}\ket{\downarrow\uparrow}$, and $P_{\downarrow\uparrow}=\cos^2{\frac{Jt}{2\hbar}}=\frac{1}{2}(1+\cos{\frac{Jt}{\hbar}})$ will thus oscillate at an angular frequency of $\frac{J}{\hbar}$. 

\begin{figure}[!htbp]
\includegraphics[width=\columnwidth]{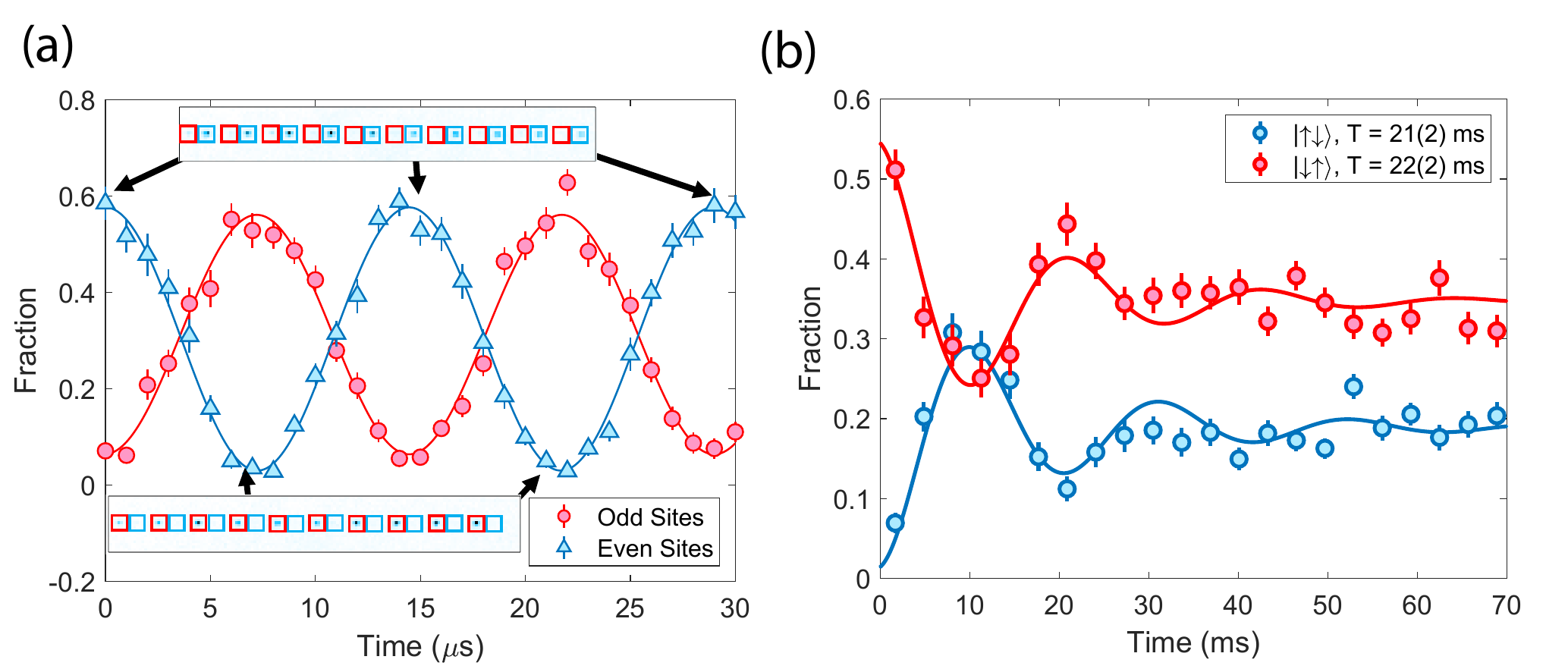}
\caption{\textbf{Single-site resolved state preparation} (a) Rabi oscillation after preparation of the molecular pair in $\ket{\downarrow\uparrow}$. (b) Dipolar spin-exchange oscillation with a initial $\ket{\uparrow\downarrow}$ state at $\theta=90\degree$. Shown here is the outcome of $\ket{\downarrow\uparrow}$ and $\ket{\uparrow\downarrow}$. }
\label{fig:dual_aod}
\end{figure}

To demonstrate individual addressing, we first prepare the molecules (both in the even and odd sites) in initial state $\ket{\downarrow\downarrow}$. We then adiabatically ramp the trap depth of the odd sites to seven times that of the even sites, so as to detune the transition of the molecules in the odd sites out of resonance. The microwave $\pi$-pulse then only transfers the molecules in the even sites from $\ket{\downarrow}$ to $\ket{\uparrow}$. This creates an anti-ferromagnetic state $\ket{\downarrow\uparrow}$. By then applying a microwave pulse with variable length of time and detecting molecules in the $\ket{\uparrow}$ state, we observe Rabi oscillations in both even and odd sites with opposite phase (Fig.~\ref{fig:dual_aod}~(a)). To observe dipolar spin-exchange, we move the tweezers to a smaller spacing ($|\vec{R}|=2\,\mu\text{m}$). As with the single AOD system, the XY8 dynamical decoupling pulses are applied and then the pairs separated for detection. The resultant outcome probabilities are shown in Fig.~\ref{fig:dual_aod}~(b), with a clear display of spin exchange.

\section{Conclusion and Outlook}
In conclusion, we observe dipolar spin-exchange interactions and create Bell state entangled pairs with single CaF molecules trapped in optical tweezers. We study the dipolar interaction and entanglement by tuning the spacing of the optical tweezers and the angle of the electric dipole quantization axis. A maximally entangled Bell state is created and its fidelity is determined using a parity oscillation measurement. Parity measurements confirm that the interaction in this system can be tuned between ferromagnetic and anti-ferromagnetic.

The coherence time of dipolar interactions and the single molecule rotational coherence time are both limited by the finite temperature of the molecules. Implementation of further cooling using other techniques, for example, Raman sideband cooling, would dramatically reduce motional dephasing~\cite{caldwell2020sideband}. The fidelity of the Bell state is also limited by the initial state preparation and detection fidelity, which can be improved by an optimized optical pumping scheme~\cite{holland2022demand}. The approach can be extended to ultracold polyatomic molecules, which have robust parity doublet states that give rise to an advantageous Stark level structure~\cite{yu2019scalable,hallas2022optical}.

\section{Acknowledgments}
\begin{acknowledgments}
This material is based upon work supported by the U.S. Department of Energy, Office of Science, National Quantum Information Science Research Centers, Quantum Systems Accelerator. Additional support is acknowledged from AFOSR, ARO and NSF. SY acknowledges support from the NSF GRFP. LA and SY acknowledge support from the HQI. EC acknowledges support from the NRF of Korea (2020R1A4A1018015, 2021M3H3A1085299, 2022M3E4A1077340).
\end{acknowledgments}

\bibliography{dipolar}

\begin{thebibliography}{55}%
\makeatletter
\providecommand \@ifxundefined [1]{%
 \@ifx{#1\undefined}
}%
\providecommand \@ifnum [1]{%
 \ifnum #1\expandafter \@firstoftwo
 \else \expandafter \@secondoftwo
 \fi
}%
\providecommand \@ifx [1]{%
 \ifx #1\expandafter \@firstoftwo
 \else \expandafter \@secondoftwo
 \fi
}%
\providecommand \natexlab [1]{#1}%
\providecommand \enquote  [1]{``#1''}%
\providecommand \bibnamefont  [1]{#1}%
\providecommand \bibfnamefont [1]{#1}%
\providecommand \citenamefont [1]{#1}%
\providecommand \href@noop [0]{\@secondoftwo}%
\providecommand \href [0]{\begingroup \@sanitize@url \@href}%
\providecommand \@href[1]{\@@startlink{#1}\@@href}%
\providecommand \@@href[1]{\endgroup#1\@@endlink}%
\providecommand \@sanitize@url [0]{\catcode `\\12\catcode `\$12\catcode
  `\&12\catcode `\#12\catcode `\^12\catcode `\_12\catcode `\%12\relax}%
\providecommand \@@startlink[1]{}%
\providecommand \@@endlink[0]{}%
\providecommand \url  [0]{\begingroup\@sanitize@url \@url }%
\providecommand \@url [1]{\endgroup\@href {#1}{\urlprefix }}%
\providecommand \urlprefix  [0]{URL }%
\providecommand \Eprint [0]{\href }%
\providecommand \doibase [0]{https://doi.org/}%
\providecommand \selectlanguage [0]{\@gobble}%
\providecommand \bibinfo  [0]{\@secondoftwo}%
\providecommand \bibfield  [0]{\@secondoftwo}%
\providecommand \translation [1]{[#1]}%
\providecommand \BibitemOpen [0]{}%
\providecommand \bibitemStop [0]{}%
\providecommand \bibitemNoStop [0]{.\EOS\space}%
\providecommand \EOS [0]{\spacefactor3000\relax}%
\providecommand \BibitemShut  [1]{\csname bibitem#1\endcsname}%
\let\auto@bib@innerbib\@empty
\bibitem [{\citenamefont {Micheli}\ \emph {et~al.}(2006)\citenamefont
  {Micheli}, \citenamefont {Brennen},\ and\ \citenamefont
  {Zoller}}]{micheli2006toolbox}%
  \BibitemOpen
  \bibfield  {author} {\bibinfo {author} {\bibfnamefont {A.}~\bibnamefont
  {Micheli}}, \bibinfo {author} {\bibfnamefont {G.}~\bibnamefont {Brennen}},\
  and\ \bibinfo {author} {\bibfnamefont {P.}~\bibnamefont {Zoller}},\
  }\bibfield  {title} {\bibinfo {title} {A toolbox for lattice-spin models with
  polar molecules},\ }\href@noop {} {\bibfield  {journal} {\bibinfo  {journal}
  {Nature Physics}\ }\textbf {\bibinfo {volume} {2}},\ \bibinfo {pages} {341}
  (\bibinfo {year} {2006})}\BibitemShut {NoStop}%
\bibitem [{\citenamefont {Gadway}\ and\ \citenamefont
  {Yan}(2016)}]{gadway2016strongly}%
  \BibitemOpen
  \bibfield  {author} {\bibinfo {author} {\bibfnamefont {B.}~\bibnamefont
  {Gadway}}\ and\ \bibinfo {author} {\bibfnamefont {B.}~\bibnamefont {Yan}},\
  }\bibfield  {title} {\bibinfo {title} {Strongly interacting ultracold polar
  molecules},\ }\href@noop {} {\bibfield  {journal} {\bibinfo  {journal}
  {Journal of Physics B: Atomic, Molecular and Optical Physics}\ }\textbf
  {\bibinfo {volume} {49}},\ \bibinfo {pages} {152002} (\bibinfo {year}
  {2016})}\BibitemShut {NoStop}%
\bibitem [{\citenamefont {Baranov}(2008)}]{baranov2008theoretical}%
  \BibitemOpen
  \bibfield  {author} {\bibinfo {author} {\bibfnamefont {M.~A.}\ \bibnamefont
  {Baranov}},\ }\bibfield  {title} {\bibinfo {title} {Theoretical progress in
  many-body physics with ultracold dipolar gases},\ }\href@noop {} {\bibfield
  {journal} {\bibinfo  {journal} {Physics Reports}\ }\textbf {\bibinfo {volume}
  {464}},\ \bibinfo {pages} {71} (\bibinfo {year} {2008})}\BibitemShut
  {NoStop}%
\bibitem [{\citenamefont {Wall}\ \emph {et~al.}(2015)\citenamefont {Wall},
  \citenamefont {Hazzard},\ and\ \citenamefont {Rey}}]{wall2015magnetism}%
  \BibitemOpen
  \bibfield  {author} {\bibinfo {author} {\bibfnamefont {M.}~\bibnamefont
  {Wall}}, \bibinfo {author} {\bibfnamefont {K.}~\bibnamefont {Hazzard}},\ and\
  \bibinfo {author} {\bibfnamefont {A.~M.}\ \bibnamefont {Rey}},\ }\bibfield
  {title} {\bibinfo {title} {Quantum magnetism with ultracold molecules},\ }in\
  \href@noop {} {\emph {\bibinfo {booktitle} {From atomic to mesoscale: The
  Role of Quantum Coherence in Systems of Various Complexities}}}\ (\bibinfo
  {publisher} {World Scientific},\ \bibinfo {year} {2015})\ pp.\ \bibinfo
  {pages} {3--37}\BibitemShut {NoStop}%
\bibitem [{\citenamefont {DeMille}(2002)}]{demille2002quantum}%
  \BibitemOpen
  \bibfield  {author} {\bibinfo {author} {\bibfnamefont {D.}~\bibnamefont
  {DeMille}},\ }\bibfield  {title} {\bibinfo {title} {Quantum computation with
  trapped polar molecules},\ }\href@noop {} {\bibfield  {journal} {\bibinfo
  {journal} {Physical Review Letters}\ }\textbf {\bibinfo {volume} {88}},\
  \bibinfo {pages} {067901} (\bibinfo {year} {2002})}\BibitemShut {NoStop}%
\bibitem [{\citenamefont {Yelin}\ \emph {et~al.}(2006)\citenamefont {Yelin},
  \citenamefont {Kirby},\ and\ \citenamefont
  {C{\^o}t{\'e}}}]{yelin2006dipolarQC}%
  \BibitemOpen
  \bibfield  {author} {\bibinfo {author} {\bibfnamefont {S.}~\bibnamefont
  {Yelin}}, \bibinfo {author} {\bibfnamefont {K.}~\bibnamefont {Kirby}},\ and\
  \bibinfo {author} {\bibfnamefont {R.}~\bibnamefont {C{\^o}t{\'e}}},\
  }\bibfield  {title} {\bibinfo {title} {Schemes for robust quantum computation
  with polar molecules},\ }\href@noop {} {\bibfield  {journal} {\bibinfo
  {journal} {Physical Review A}\ }\textbf {\bibinfo {volume} {74}},\ \bibinfo
  {pages} {050301} (\bibinfo {year} {2006})}\BibitemShut {NoStop}%
\bibitem [{\citenamefont {Karra}\ \emph {et~al.}(2016)\citenamefont {Karra},
  \citenamefont {Sharma}, \citenamefont {Friedrich}, \citenamefont {Kais},\
  and\ \citenamefont {Herschbach}}]{karra2016paramagnetic}%
  \BibitemOpen
  \bibfield  {author} {\bibinfo {author} {\bibfnamefont {M.}~\bibnamefont
  {Karra}}, \bibinfo {author} {\bibfnamefont {K.}~\bibnamefont {Sharma}},
  \bibinfo {author} {\bibfnamefont {B.}~\bibnamefont {Friedrich}}, \bibinfo
  {author} {\bibfnamefont {S.}~\bibnamefont {Kais}},\ and\ \bibinfo {author}
  {\bibfnamefont {D.}~\bibnamefont {Herschbach}},\ }\bibfield  {title}
  {\bibinfo {title} {Prospects for quantum computing with an array of ultracold
  polar paramagnetic molecules},\ }\href@noop {} {\bibfield  {journal}
  {\bibinfo  {journal} {The Journal of chemical physics}\ }\textbf {\bibinfo
  {volume} {144}},\ \bibinfo {pages} {094301} (\bibinfo {year}
  {2016})}\BibitemShut {NoStop}%
\bibitem [{\citenamefont {Sawant}\ \emph {et~al.}(2020)\citenamefont {Sawant},
  \citenamefont {Blackmore}, \citenamefont {Gregory}, \citenamefont
  {Mur-Petit}, \citenamefont {Jaksch}, \citenamefont {Aldegunde}, \citenamefont
  {Hutson}, \citenamefont {Tarbutt},\ and\ \citenamefont
  {Cornish}}]{sawant2020qudits}%
  \BibitemOpen
  \bibfield  {author} {\bibinfo {author} {\bibfnamefont {R.}~\bibnamefont
  {Sawant}}, \bibinfo {author} {\bibfnamefont {J.~A.}\ \bibnamefont
  {Blackmore}}, \bibinfo {author} {\bibfnamefont {P.~D.}\ \bibnamefont
  {Gregory}}, \bibinfo {author} {\bibfnamefont {J.}~\bibnamefont {Mur-Petit}},
  \bibinfo {author} {\bibfnamefont {D.}~\bibnamefont {Jaksch}}, \bibinfo
  {author} {\bibfnamefont {J.}~\bibnamefont {Aldegunde}}, \bibinfo {author}
  {\bibfnamefont {J.~M.}\ \bibnamefont {Hutson}}, \bibinfo {author}
  {\bibfnamefont {M.}~\bibnamefont {Tarbutt}},\ and\ \bibinfo {author}
  {\bibfnamefont {S.~L.}\ \bibnamefont {Cornish}},\ }\bibfield  {title}
  {\bibinfo {title} {Ultracold polar molecules as qudits},\ }\href@noop {}
  {\bibfield  {journal} {\bibinfo  {journal} {New Journal of Physics}\ }\textbf
  {\bibinfo {volume} {22}},\ \bibinfo {pages} {013027} (\bibinfo {year}
  {2020})}\BibitemShut {NoStop}%
\bibitem [{\citenamefont {Ni}\ \emph {et~al.}(2018)\citenamefont {Ni},
  \citenamefont {Rosenband},\ and\ \citenamefont {Grimes}}]{ni2018dipolar}%
  \BibitemOpen
  \bibfield  {author} {\bibinfo {author} {\bibfnamefont {K.-K.}\ \bibnamefont
  {Ni}}, \bibinfo {author} {\bibfnamefont {T.}~\bibnamefont {Rosenband}},\ and\
  \bibinfo {author} {\bibfnamefont {D.~D.}\ \bibnamefont {Grimes}},\ }\bibfield
   {title} {\bibinfo {title} {Dipolar exchange quantum logic gate with polar
  molecules},\ }\href@noop {} {\bibfield  {journal} {\bibinfo  {journal}
  {Chemical science}\ }\textbf {\bibinfo {volume} {9}},\ \bibinfo {pages}
  {6830} (\bibinfo {year} {2018})}\BibitemShut {NoStop}%
\bibitem [{\citenamefont {Carr}\ \emph {et~al.}(2009)\citenamefont {Carr},
  \citenamefont {DeMille}, \citenamefont {Krems},\ and\ \citenamefont
  {Ye}}]{carr2009coldmolecules}%
  \BibitemOpen
  \bibfield  {author} {\bibinfo {author} {\bibfnamefont {L.~D.}\ \bibnamefont
  {Carr}}, \bibinfo {author} {\bibfnamefont {D.}~\bibnamefont {DeMille}},
  \bibinfo {author} {\bibfnamefont {R.~V.}\ \bibnamefont {Krems}},\ and\
  \bibinfo {author} {\bibfnamefont {J.}~\bibnamefont {Ye}},\ }\bibfield
  {title} {\bibinfo {title} {Cold and ultracold molecules: science, technology
  and applications},\ }\href@noop {} {\bibfield  {journal} {\bibinfo  {journal}
  {New Journal of Physics}\ }\textbf {\bibinfo {volume} {11}},\ \bibinfo
  {pages} {055049} (\bibinfo {year} {2009})}\BibitemShut {NoStop}%
\bibitem [{\citenamefont {Barry}\ \emph {et~al.}(2014)\citenamefont {Barry},
  \citenamefont {McCarron}, \citenamefont {Norrgard}, \citenamefont
  {Steinecker},\ and\ \citenamefont {DeMille}}]{barry2014magneto}%
  \BibitemOpen
  \bibfield  {author} {\bibinfo {author} {\bibfnamefont {J.}~\bibnamefont
  {Barry}}, \bibinfo {author} {\bibfnamefont {D.}~\bibnamefont {McCarron}},
  \bibinfo {author} {\bibfnamefont {E.}~\bibnamefont {Norrgard}}, \bibinfo
  {author} {\bibfnamefont {M.}~\bibnamefont {Steinecker}},\ and\ \bibinfo
  {author} {\bibfnamefont {D.}~\bibnamefont {DeMille}},\ }\bibfield  {title}
  {\bibinfo {title} {Magneto-optical trapping of a diatomic molecule},\
  }\href@noop {} {\bibfield  {journal} {\bibinfo  {journal} {Nature}\ }\textbf
  {\bibinfo {volume} {512}},\ \bibinfo {pages} {286} (\bibinfo {year}
  {2014})}\BibitemShut {NoStop}%
\bibitem [{\citenamefont {Anderegg}\ \emph {et~al.}(2017)\citenamefont
  {Anderegg}, \citenamefont {Augenbraun}, \citenamefont {Chae}, \citenamefont
  {Hemmerling}, \citenamefont {Hutzler}, \citenamefont {Ravi}, \citenamefont
  {Collopy}, \citenamefont {Ye}, \citenamefont {Ketterle},\ and\ \citenamefont
  {Doyle}}]{anderegg2017radio}%
  \BibitemOpen
  \bibfield  {author} {\bibinfo {author} {\bibfnamefont {L.}~\bibnamefont
  {Anderegg}}, \bibinfo {author} {\bibfnamefont {B.~L.}\ \bibnamefont
  {Augenbraun}}, \bibinfo {author} {\bibfnamefont {E.}~\bibnamefont {Chae}},
  \bibinfo {author} {\bibfnamefont {B.}~\bibnamefont {Hemmerling}}, \bibinfo
  {author} {\bibfnamefont {N.~R.}\ \bibnamefont {Hutzler}}, \bibinfo {author}
  {\bibfnamefont {A.}~\bibnamefont {Ravi}}, \bibinfo {author} {\bibfnamefont
  {A.}~\bibnamefont {Collopy}}, \bibinfo {author} {\bibfnamefont
  {J.}~\bibnamefont {Ye}}, \bibinfo {author} {\bibfnamefont {W.}~\bibnamefont
  {Ketterle}},\ and\ \bibinfo {author} {\bibfnamefont {J.~M.}\ \bibnamefont
  {Doyle}},\ }\bibfield  {title} {\bibinfo {title} {Radio frequency
  magneto-optical trapping of $\text{CaF}$ with high density},\ }\href@noop {}
  {\bibfield  {journal} {\bibinfo  {journal} {Physical Review Letters}\
  }\textbf {\bibinfo {volume} {119}},\ \bibinfo {pages} {103201} (\bibinfo
  {year} {2017})}\BibitemShut {NoStop}%
\bibitem [{\citenamefont {Truppe}\ \emph {et~al.}(2017)\citenamefont {Truppe},
  \citenamefont {Williams}, \citenamefont {Hambach}, \citenamefont {Caldwell},
  \citenamefont {Fitch}, \citenamefont {Hinds}, \citenamefont {Sauer},\ and\
  \citenamefont {Tarbutt}}]{truppe2017molecules}%
  \BibitemOpen
  \bibfield  {author} {\bibinfo {author} {\bibfnamefont {S.}~\bibnamefont
  {Truppe}}, \bibinfo {author} {\bibfnamefont {H.}~\bibnamefont {Williams}},
  \bibinfo {author} {\bibfnamefont {M.}~\bibnamefont {Hambach}}, \bibinfo
  {author} {\bibfnamefont {L.}~\bibnamefont {Caldwell}}, \bibinfo {author}
  {\bibfnamefont {N.}~\bibnamefont {Fitch}}, \bibinfo {author} {\bibfnamefont
  {E.}~\bibnamefont {Hinds}}, \bibinfo {author} {\bibfnamefont
  {B.}~\bibnamefont {Sauer}},\ and\ \bibinfo {author} {\bibfnamefont
  {M.}~\bibnamefont {Tarbutt}},\ }\bibfield  {title} {\bibinfo {title}
  {Molecules cooled below the doppler limit},\ }\href@noop {} {\bibfield
  {journal} {\bibinfo  {journal} {Nature Physics}\ }\textbf {\bibinfo {volume}
  {13}},\ \bibinfo {pages} {1173} (\bibinfo {year} {2017})}\BibitemShut
  {NoStop}%
\bibitem [{\citenamefont {Collopy}\ \emph {et~al.}(2018)\citenamefont
  {Collopy}, \citenamefont {Ding}, \citenamefont {Wu}, \citenamefont
  {Finneran}, \citenamefont {Anderegg}, \citenamefont {Augenbraun},
  \citenamefont {Doyle},\ and\ \citenamefont {Ye}}]{collopy20183d}%
  \BibitemOpen
  \bibfield  {author} {\bibinfo {author} {\bibfnamefont {A.~L.}\ \bibnamefont
  {Collopy}}, \bibinfo {author} {\bibfnamefont {S.}~\bibnamefont {Ding}},
  \bibinfo {author} {\bibfnamefont {Y.}~\bibnamefont {Wu}}, \bibinfo {author}
  {\bibfnamefont {I.~A.}\ \bibnamefont {Finneran}}, \bibinfo {author}
  {\bibfnamefont {L.}~\bibnamefont {Anderegg}}, \bibinfo {author}
  {\bibfnamefont {B.~L.}\ \bibnamefont {Augenbraun}}, \bibinfo {author}
  {\bibfnamefont {J.~M.}\ \bibnamefont {Doyle}},\ and\ \bibinfo {author}
  {\bibfnamefont {J.}~\bibnamefont {Ye}},\ }\bibfield  {title} {\bibinfo
  {title} {3d magneto-optical trap of yttrium monoxide},\ }\href@noop {}
  {\bibfield  {journal} {\bibinfo  {journal} {Physical Review Letters}\
  }\textbf {\bibinfo {volume} {121}},\ \bibinfo {pages} {213201} (\bibinfo
  {year} {2018})}\BibitemShut {NoStop}%
\bibitem [{\citenamefont {Vilas}\ \emph {et~al.}(2022)\citenamefont {Vilas},
  \citenamefont {Hallas}, \citenamefont {Anderegg}, \citenamefont {Robichaud},
  \citenamefont {Winnicki}, \citenamefont {Mitra},\ and\ \citenamefont
  {Doyle}}]{vilas2022CaOHMOT}%
  \BibitemOpen
  \bibfield  {author} {\bibinfo {author} {\bibfnamefont {N.~B.}\ \bibnamefont
  {Vilas}}, \bibinfo {author} {\bibfnamefont {C.}~\bibnamefont {Hallas}},
  \bibinfo {author} {\bibfnamefont {L.}~\bibnamefont {Anderegg}}, \bibinfo
  {author} {\bibfnamefont {P.}~\bibnamefont {Robichaud}}, \bibinfo {author}
  {\bibfnamefont {A.}~\bibnamefont {Winnicki}}, \bibinfo {author}
  {\bibfnamefont {D.}~\bibnamefont {Mitra}},\ and\ \bibinfo {author}
  {\bibfnamefont {J.~M.}\ \bibnamefont {Doyle}},\ }\bibfield  {title} {\bibinfo
  {title} {Magneto-optical trapping and sub-doppler cooling of a polyatomic
  molecule},\ }\href@noop {} {\bibfield  {journal} {\bibinfo  {journal}
  {Nature}\ }\textbf {\bibinfo {volume} {606}},\ \bibinfo {pages} {70}
  (\bibinfo {year} {2022})}\BibitemShut {NoStop}%
\bibitem [{\citenamefont {Ni}\ \emph {et~al.}(2008)\citenamefont {Ni},
  \citenamefont {Ospelkaus}, \citenamefont {De~Miranda}, \citenamefont {Pe'Er},
  \citenamefont {Neyenhuis}, \citenamefont {Zirbel}, \citenamefont
  {Kotochigova}, \citenamefont {Julienne}, \citenamefont {Jin},\ and\
  \citenamefont {Ye}}]{ni2008high}%
  \BibitemOpen
  \bibfield  {author} {\bibinfo {author} {\bibfnamefont {K.-K.}\ \bibnamefont
  {Ni}}, \bibinfo {author} {\bibfnamefont {S.}~\bibnamefont {Ospelkaus}},
  \bibinfo {author} {\bibfnamefont {M.}~\bibnamefont {De~Miranda}}, \bibinfo
  {author} {\bibfnamefont {A.}~\bibnamefont {Pe'Er}}, \bibinfo {author}
  {\bibfnamefont {B.}~\bibnamefont {Neyenhuis}}, \bibinfo {author}
  {\bibfnamefont {J.}~\bibnamefont {Zirbel}}, \bibinfo {author} {\bibfnamefont
  {S.}~\bibnamefont {Kotochigova}}, \bibinfo {author} {\bibfnamefont
  {P.}~\bibnamefont {Julienne}}, \bibinfo {author} {\bibfnamefont
  {D.}~\bibnamefont {Jin}},\ and\ \bibinfo {author} {\bibfnamefont
  {J.}~\bibnamefont {Ye}},\ }\bibfield  {title} {\bibinfo {title} {A high
  phase-space-density gas of polar molecules},\ }\href@noop {} {\bibfield
  {journal} {\bibinfo  {journal} {science}\ }\textbf {\bibinfo {volume}
  {322}},\ \bibinfo {pages} {231} (\bibinfo {year} {2008})}\BibitemShut
  {NoStop}%
\bibitem [{\citenamefont {Molony}\ \emph {et~al.}(2014)\citenamefont {Molony},
  \citenamefont {Gregory}, \citenamefont {Ji}, \citenamefont {Lu},
  \citenamefont {K{\"o}ppinger}, \citenamefont {Le~Sueur}, \citenamefont
  {Blackley}, \citenamefont {Hutson},\ and\ \citenamefont
  {Cornish}}]{molony2014creation}%
  \BibitemOpen
  \bibfield  {author} {\bibinfo {author} {\bibfnamefont {P.~K.}\ \bibnamefont
  {Molony}}, \bibinfo {author} {\bibfnamefont {P.~D.}\ \bibnamefont {Gregory}},
  \bibinfo {author} {\bibfnamefont {Z.}~\bibnamefont {Ji}}, \bibinfo {author}
  {\bibfnamefont {B.}~\bibnamefont {Lu}}, \bibinfo {author} {\bibfnamefont
  {M.~P.}\ \bibnamefont {K{\"o}ppinger}}, \bibinfo {author} {\bibfnamefont
  {C.~R.}\ \bibnamefont {Le~Sueur}}, \bibinfo {author} {\bibfnamefont {C.~L.}\
  \bibnamefont {Blackley}}, \bibinfo {author} {\bibfnamefont {J.~M.}\
  \bibnamefont {Hutson}},\ and\ \bibinfo {author} {\bibfnamefont {S.~L.}\
  \bibnamefont {Cornish}},\ }\bibfield  {title} {\bibinfo {title} {Creation of
  ultracold $^{Rb}\text{Rb} ^{133}\text{Cs}$ molecules in the rovibrational
  ground state},\ }\href@noop {} {\bibfield  {journal} {\bibinfo  {journal}
  {Physical Review Letters}\ }\textbf {\bibinfo {volume} {113}},\ \bibinfo
  {pages} {255301} (\bibinfo {year} {2014})}\BibitemShut {NoStop}%
\bibitem [{\citenamefont {Takekoshi}\ \emph {et~al.}(2014)\citenamefont
  {Takekoshi}, \citenamefont {Reichs{\"o}llner}, \citenamefont {Schindewolf},
  \citenamefont {Hutson}, \citenamefont {Le~Sueur}, \citenamefont {Dulieu},
  \citenamefont {Ferlaino}, \citenamefont {Grimm},\ and\ \citenamefont
  {N{\"a}gerl}}]{takekoshi2014ultracold}%
  \BibitemOpen
  \bibfield  {author} {\bibinfo {author} {\bibfnamefont {T.}~\bibnamefont
  {Takekoshi}}, \bibinfo {author} {\bibfnamefont {L.}~\bibnamefont
  {Reichs{\"o}llner}}, \bibinfo {author} {\bibfnamefont {A.}~\bibnamefont
  {Schindewolf}}, \bibinfo {author} {\bibfnamefont {J.~M.}\ \bibnamefont
  {Hutson}}, \bibinfo {author} {\bibfnamefont {C.~R.}\ \bibnamefont
  {Le~Sueur}}, \bibinfo {author} {\bibfnamefont {O.}~\bibnamefont {Dulieu}},
  \bibinfo {author} {\bibfnamefont {F.}~\bibnamefont {Ferlaino}}, \bibinfo
  {author} {\bibfnamefont {R.}~\bibnamefont {Grimm}},\ and\ \bibinfo {author}
  {\bibfnamefont {H.-C.}\ \bibnamefont {N{\"a}gerl}},\ }\bibfield  {title}
  {\bibinfo {title} {Ultracold dense samples of dipolar rbcs molecules in the
  rovibrational and hyperfine ground state},\ }\href@noop {} {\bibfield
  {journal} {\bibinfo  {journal} {Physical review letters}\ }\textbf {\bibinfo
  {volume} {113}},\ \bibinfo {pages} {205301} (\bibinfo {year}
  {2014})}\BibitemShut {NoStop}%
\bibitem [{\citenamefont {Park}\ \emph {et~al.}(2015)\citenamefont {Park},
  \citenamefont {Will},\ and\ \citenamefont {Zwierlein}}]{park2015ultracold}%
  \BibitemOpen
  \bibfield  {author} {\bibinfo {author} {\bibfnamefont {J.~W.}\ \bibnamefont
  {Park}}, \bibinfo {author} {\bibfnamefont {S.~A.}\ \bibnamefont {Will}},\
  and\ \bibinfo {author} {\bibfnamefont {M.~W.}\ \bibnamefont {Zwierlein}},\
  }\bibfield  {title} {\bibinfo {title} {Ultracold dipolar gas of fermionic
  $^{23}\text{Na} ^{40}\text{K}$ molecules in their absolute ground state},\
  }\href@noop {} {\bibfield  {journal} {\bibinfo  {journal} {Physical Review
  Letters}\ }\textbf {\bibinfo {volume} {114}},\ \bibinfo {pages} {205302}
  (\bibinfo {year} {2015})}\BibitemShut {NoStop}%
\bibitem [{\citenamefont {Guo}\ \emph {et~al.}(2016)\citenamefont {Guo},
  \citenamefont {Zhu}, \citenamefont {Lu}, \citenamefont {Ye}, \citenamefont
  {Wang}, \citenamefont {Vexiau}, \citenamefont {Bouloufa-Maafa}, \citenamefont
  {Qu{\'e}m{\'e}ner}, \citenamefont {Dulieu},\ and\ \citenamefont
  {Wang}}]{guo2016creation}%
  \BibitemOpen
  \bibfield  {author} {\bibinfo {author} {\bibfnamefont {M.}~\bibnamefont
  {Guo}}, \bibinfo {author} {\bibfnamefont {B.}~\bibnamefont {Zhu}}, \bibinfo
  {author} {\bibfnamefont {B.}~\bibnamefont {Lu}}, \bibinfo {author}
  {\bibfnamefont {X.}~\bibnamefont {Ye}}, \bibinfo {author} {\bibfnamefont
  {F.}~\bibnamefont {Wang}}, \bibinfo {author} {\bibfnamefont {R.}~\bibnamefont
  {Vexiau}}, \bibinfo {author} {\bibfnamefont {N.}~\bibnamefont
  {Bouloufa-Maafa}}, \bibinfo {author} {\bibfnamefont {G.}~\bibnamefont
  {Qu{\'e}m{\'e}ner}}, \bibinfo {author} {\bibfnamefont {O.}~\bibnamefont
  {Dulieu}},\ and\ \bibinfo {author} {\bibfnamefont {D.}~\bibnamefont {Wang}},\
  }\bibfield  {title} {\bibinfo {title} {Creation of an ultracold gas of
  ground-state dipolar $^{23}\text{Na} ^{87}\text{Rb}$ molecules},\ }\href@noop
  {} {\bibfield  {journal} {\bibinfo  {journal} {Physical Review Letters}\
  }\textbf {\bibinfo {volume} {116}},\ \bibinfo {pages} {205303} (\bibinfo
  {year} {2016})}\BibitemShut {NoStop}%
\bibitem [{\citenamefont {Rvachov}\ \emph {et~al.}(2017)\citenamefont
  {Rvachov}, \citenamefont {Son}, \citenamefont {Sommer}, \citenamefont
  {Ebadi}, \citenamefont {Park}, \citenamefont {Zwierlein}, \citenamefont
  {Ketterle},\ and\ \citenamefont {Jamison}}]{rvachov2017long}%
  \BibitemOpen
  \bibfield  {author} {\bibinfo {author} {\bibfnamefont {T.~M.}\ \bibnamefont
  {Rvachov}}, \bibinfo {author} {\bibfnamefont {H.}~\bibnamefont {Son}},
  \bibinfo {author} {\bibfnamefont {A.~T.}\ \bibnamefont {Sommer}}, \bibinfo
  {author} {\bibfnamefont {S.}~\bibnamefont {Ebadi}}, \bibinfo {author}
  {\bibfnamefont {J.~J.}\ \bibnamefont {Park}}, \bibinfo {author}
  {\bibfnamefont {M.~W.}\ \bibnamefont {Zwierlein}}, \bibinfo {author}
  {\bibfnamefont {W.}~\bibnamefont {Ketterle}},\ and\ \bibinfo {author}
  {\bibfnamefont {A.~O.}\ \bibnamefont {Jamison}},\ }\bibfield  {title}
  {\bibinfo {title} {Long-lived ultracold molecules with electric and magnetic
  dipole moments},\ }\href@noop {} {\bibfield  {journal} {\bibinfo  {journal}
  {Physical Review Letters}\ }\textbf {\bibinfo {volume} {119}},\ \bibinfo
  {pages} {143001} (\bibinfo {year} {2017})}\BibitemShut {NoStop}%
\bibitem [{\citenamefont {Liu}\ \emph {et~al.}(2018)\citenamefont {Liu},
  \citenamefont {Hood}, \citenamefont {Yu}, \citenamefont {Zhang},
  \citenamefont {Hutzler}, \citenamefont {Rosenband},\ and\ \citenamefont
  {Ni}}]{liu2018building}%
  \BibitemOpen
  \bibfield  {author} {\bibinfo {author} {\bibfnamefont {L.}~\bibnamefont
  {Liu}}, \bibinfo {author} {\bibfnamefont {J.}~\bibnamefont {Hood}}, \bibinfo
  {author} {\bibfnamefont {Y.}~\bibnamefont {Yu}}, \bibinfo {author}
  {\bibfnamefont {J.}~\bibnamefont {Zhang}}, \bibinfo {author} {\bibfnamefont
  {N.}~\bibnamefont {Hutzler}}, \bibinfo {author} {\bibfnamefont
  {T.}~\bibnamefont {Rosenband}},\ and\ \bibinfo {author} {\bibfnamefont
  {K.-K.}\ \bibnamefont {Ni}},\ }\bibfield  {title} {\bibinfo {title} {Building
  one molecule from a reservoir of two atoms},\ }\href@noop {} {\bibfield
  {journal} {\bibinfo  {journal} {Science}\ }\textbf {\bibinfo {volume}
  {360}},\ \bibinfo {pages} {900} (\bibinfo {year} {2018})}\BibitemShut
  {NoStop}%
\bibitem [{\citenamefont {Yan}\ \emph {et~al.}(2013)\citenamefont {Yan},
  \citenamefont {Moses}, \citenamefont {Gadway}, \citenamefont {Covey},
  \citenamefont {Hazzard}, \citenamefont {Rey}, \citenamefont {Jin},\ and\
  \citenamefont {Ye}}]{yan2013observation}%
  \BibitemOpen
  \bibfield  {author} {\bibinfo {author} {\bibfnamefont {B.}~\bibnamefont
  {Yan}}, \bibinfo {author} {\bibfnamefont {S.~A.}\ \bibnamefont {Moses}},
  \bibinfo {author} {\bibfnamefont {B.}~\bibnamefont {Gadway}}, \bibinfo
  {author} {\bibfnamefont {J.~P.}\ \bibnamefont {Covey}}, \bibinfo {author}
  {\bibfnamefont {K.~R.}\ \bibnamefont {Hazzard}}, \bibinfo {author}
  {\bibfnamefont {A.~M.}\ \bibnamefont {Rey}}, \bibinfo {author} {\bibfnamefont
  {D.~S.}\ \bibnamefont {Jin}},\ and\ \bibinfo {author} {\bibfnamefont
  {J.}~\bibnamefont {Ye}},\ }\bibfield  {title} {\bibinfo {title} {Observation
  of dipolar spin-exchange interactions with lattice-confined polar
  molecules},\ }\href@noop {} {\bibfield  {journal} {\bibinfo  {journal}
  {Nature}\ }\textbf {\bibinfo {volume} {501}},\ \bibinfo {pages} {521}
  (\bibinfo {year} {2013})}\BibitemShut {NoStop}%
\bibitem [{\citenamefont {See{\ss}elberg}\ \emph {et~al.}(2018)\citenamefont
  {See{\ss}elberg}, \citenamefont {Luo}, \citenamefont {Li}, \citenamefont
  {Bause}, \citenamefont {Kotochigova}, \citenamefont {Bloch},\ and\
  \citenamefont {Gohle}}]{seesselberg2018extending}%
  \BibitemOpen
  \bibfield  {author} {\bibinfo {author} {\bibfnamefont {F.}~\bibnamefont
  {See{\ss}elberg}}, \bibinfo {author} {\bibfnamefont {X.-Y.}\ \bibnamefont
  {Luo}}, \bibinfo {author} {\bibfnamefont {M.}~\bibnamefont {Li}}, \bibinfo
  {author} {\bibfnamefont {R.}~\bibnamefont {Bause}}, \bibinfo {author}
  {\bibfnamefont {S.}~\bibnamefont {Kotochigova}}, \bibinfo {author}
  {\bibfnamefont {I.}~\bibnamefont {Bloch}},\ and\ \bibinfo {author}
  {\bibfnamefont {C.}~\bibnamefont {Gohle}},\ }\bibfield  {title} {\bibinfo
  {title} {Extending rotational coherence of interacting polar molecules in a
  spin-decoupled magic trap},\ }\href@noop {} {\bibfield  {journal} {\bibinfo
  {journal} {Physical Review Letters}\ }\textbf {\bibinfo {volume} {121}},\
  \bibinfo {pages} {253401} (\bibinfo {year} {2018})}\BibitemShut {NoStop}%
\bibitem [{\citenamefont {Williams}\ \emph {et~al.}(2018)\citenamefont
  {Williams}, \citenamefont {Caldwell}, \citenamefont {Fitch}, \citenamefont
  {Truppe}, \citenamefont {Rodewald}, \citenamefont {Hinds}, \citenamefont
  {Sauer},\ and\ \citenamefont {Tarbutt}}]{williams2018magnetic}%
  \BibitemOpen
  \bibfield  {author} {\bibinfo {author} {\bibfnamefont {H.}~\bibnamefont
  {Williams}}, \bibinfo {author} {\bibfnamefont {L.}~\bibnamefont {Caldwell}},
  \bibinfo {author} {\bibfnamefont {N.}~\bibnamefont {Fitch}}, \bibinfo
  {author} {\bibfnamefont {S.}~\bibnamefont {Truppe}}, \bibinfo {author}
  {\bibfnamefont {J.}~\bibnamefont {Rodewald}}, \bibinfo {author}
  {\bibfnamefont {E.}~\bibnamefont {Hinds}}, \bibinfo {author} {\bibfnamefont
  {B.}~\bibnamefont {Sauer}},\ and\ \bibinfo {author} {\bibfnamefont
  {M.}~\bibnamefont {Tarbutt}},\ }\bibfield  {title} {\bibinfo {title}
  {Magnetic trapping and coherent control of laser-cooled molecules},\
  }\href@noop {} {\bibfield  {journal} {\bibinfo  {journal} {Physical Review
  Letters}\ }\textbf {\bibinfo {volume} {120}},\ \bibinfo {pages} {163201}
  (\bibinfo {year} {2018})}\BibitemShut {NoStop}%
\bibitem [{\citenamefont {Li}\ \emph {et~al.}(2022)\citenamefont {Li},
  \citenamefont {Matsuda}, \citenamefont {Miller}, \citenamefont {Carroll},
  \citenamefont {Tobias}, \citenamefont {Higgins},\ and\ \citenamefont
  {Ye}}]{li2022tunable}%
  \BibitemOpen
  \bibfield  {author} {\bibinfo {author} {\bibfnamefont {J.-R.}\ \bibnamefont
  {Li}}, \bibinfo {author} {\bibfnamefont {K.}~\bibnamefont {Matsuda}},
  \bibinfo {author} {\bibfnamefont {C.}~\bibnamefont {Miller}}, \bibinfo
  {author} {\bibfnamefont {A.~N.}\ \bibnamefont {Carroll}}, \bibinfo {author}
  {\bibfnamefont {W.~G.}\ \bibnamefont {Tobias}}, \bibinfo {author}
  {\bibfnamefont {J.~S.}\ \bibnamefont {Higgins}},\ and\ \bibinfo {author}
  {\bibfnamefont {J.}~\bibnamefont {Ye}},\ }\bibfield  {title} {\bibinfo
  {title} {Tunable itinerant spin dynamics with polar molecules},\ }\href@noop
  {} {\bibfield  {journal} {\bibinfo  {journal} {arXiv preprint
  arXiv:2208.02216}\ } (\bibinfo {year} {2022})}\BibitemShut {NoStop}%
\bibitem [{\citenamefont {Burchesky}\ \emph {et~al.}(2021)\citenamefont
  {Burchesky}, \citenamefont {Anderegg}, \citenamefont {Bao}, \citenamefont
  {Yu}, \citenamefont {Chae}, \citenamefont {Ketterle}, \citenamefont {Ni},\
  and\ \citenamefont {Doyle}}]{burchesky2021rotational}%
  \BibitemOpen
  \bibfield  {author} {\bibinfo {author} {\bibfnamefont {S.}~\bibnamefont
  {Burchesky}}, \bibinfo {author} {\bibfnamefont {L.}~\bibnamefont {Anderegg}},
  \bibinfo {author} {\bibfnamefont {Y.}~\bibnamefont {Bao}}, \bibinfo {author}
  {\bibfnamefont {S.~S.}\ \bibnamefont {Yu}}, \bibinfo {author} {\bibfnamefont
  {E.}~\bibnamefont {Chae}}, \bibinfo {author} {\bibfnamefont {W.}~\bibnamefont
  {Ketterle}}, \bibinfo {author} {\bibfnamefont {K.-K.}\ \bibnamefont {Ni}},\
  and\ \bibinfo {author} {\bibfnamefont {J.~M.}\ \bibnamefont {Doyle}},\
  }\bibfield  {title} {\bibinfo {title} {Rotational coherence times of polar
  molecules in optical tweezers},\ }\href@noop {} {\bibfield  {journal}
  {\bibinfo  {journal} {Physical Review Letters}\ }\textbf {\bibinfo {volume}
  {127}},\ \bibinfo {pages} {123202} (\bibinfo {year} {2021})}\BibitemShut
  {NoStop}%
\bibitem [{\citenamefont {Cairncross}\ \emph {et~al.}(2021)\citenamefont
  {Cairncross}, \citenamefont {Zhang}, \citenamefont {Picard}, \citenamefont
  {Yu}, \citenamefont {Wang},\ and\ \citenamefont
  {Ni}}]{cairncross2021assembly}%
  \BibitemOpen
  \bibfield  {author} {\bibinfo {author} {\bibfnamefont {W.~B.}\ \bibnamefont
  {Cairncross}}, \bibinfo {author} {\bibfnamefont {J.~T.}\ \bibnamefont
  {Zhang}}, \bibinfo {author} {\bibfnamefont {L.~R.}\ \bibnamefont {Picard}},
  \bibinfo {author} {\bibfnamefont {Y.}~\bibnamefont {Yu}}, \bibinfo {author}
  {\bibfnamefont {K.}~\bibnamefont {Wang}},\ and\ \bibinfo {author}
  {\bibfnamefont {K.-K.}\ \bibnamefont {Ni}},\ }\bibfield  {title} {\bibinfo
  {title} {Assembly of a rovibrational ground state molecule in an optical
  tweezer},\ }\href@noop {} {\bibfield  {journal} {\bibinfo  {journal}
  {Physical Review Letters}\ }\textbf {\bibinfo {volume} {126}},\ \bibinfo
  {pages} {123402} (\bibinfo {year} {2021})}\BibitemShut {NoStop}%
\bibitem [{\citenamefont {Christakis}\ \emph {et~al.}(2022)\citenamefont
  {Christakis}, \citenamefont {Rosenberg}, \citenamefont {Raj}, \citenamefont
  {Chi}, \citenamefont {Morningstar}, \citenamefont {Huse}, \citenamefont
  {Yan},\ and\ \citenamefont {Bakr}}]{christakis2022probing}%
  \BibitemOpen
  \bibfield  {author} {\bibinfo {author} {\bibfnamefont {L.}~\bibnamefont
  {Christakis}}, \bibinfo {author} {\bibfnamefont {J.~S.}\ \bibnamefont
  {Rosenberg}}, \bibinfo {author} {\bibfnamefont {R.}~\bibnamefont {Raj}},
  \bibinfo {author} {\bibfnamefont {S.}~\bibnamefont {Chi}}, \bibinfo {author}
  {\bibfnamefont {A.}~\bibnamefont {Morningstar}}, \bibinfo {author}
  {\bibfnamefont {D.~A.}\ \bibnamefont {Huse}}, \bibinfo {author}
  {\bibfnamefont {Z.~Z.}\ \bibnamefont {Yan}},\ and\ \bibinfo {author}
  {\bibfnamefont {W.~S.}\ \bibnamefont {Bakr}},\ }\bibfield  {title} {\bibinfo
  {title} {Probing site-resolved correlations in a spin system of ultracold
  molecules},\ }\href@noop {} {\bibfield  {journal} {\bibinfo  {journal} {arXiv
  preprint arXiv:2207.09328}\ } (\bibinfo {year} {2022})}\BibitemShut {NoStop}%
\bibitem [{\citenamefont {Endres}\ \emph {et~al.}(2016)\citenamefont {Endres},
  \citenamefont {Bernien}, \citenamefont {Keesling}, \citenamefont {Levine},
  \citenamefont {Anschuetz}, \citenamefont {Krajenbrink}, \citenamefont
  {Senko}, \citenamefont {Vuletic}, \citenamefont {Greiner},\ and\
  \citenamefont {Lukin}}]{endres2016atom}%
  \BibitemOpen
  \bibfield  {author} {\bibinfo {author} {\bibfnamefont {M.}~\bibnamefont
  {Endres}}, \bibinfo {author} {\bibfnamefont {H.}~\bibnamefont {Bernien}},
  \bibinfo {author} {\bibfnamefont {A.}~\bibnamefont {Keesling}}, \bibinfo
  {author} {\bibfnamefont {H.}~\bibnamefont {Levine}}, \bibinfo {author}
  {\bibfnamefont {E.~R.}\ \bibnamefont {Anschuetz}}, \bibinfo {author}
  {\bibfnamefont {A.}~\bibnamefont {Krajenbrink}}, \bibinfo {author}
  {\bibfnamefont {C.}~\bibnamefont {Senko}}, \bibinfo {author} {\bibfnamefont
  {V.}~\bibnamefont {Vuletic}}, \bibinfo {author} {\bibfnamefont
  {M.}~\bibnamefont {Greiner}},\ and\ \bibinfo {author} {\bibfnamefont {M.~D.}\
  \bibnamefont {Lukin}},\ }\bibfield  {title} {\bibinfo {title} {Atom-by-atom
  assembly of defect-free one-dimensional cold atom arrays},\ }\href@noop {}
  {\bibfield  {journal} {\bibinfo  {journal} {Science}\ }\textbf {\bibinfo
  {volume} {354}},\ \bibinfo {pages} {1024} (\bibinfo {year}
  {2016})}\BibitemShut {NoStop}%
\bibitem [{\citenamefont {Barredo}\ \emph {et~al.}(2016)\citenamefont
  {Barredo}, \citenamefont {De~L{\'e}s{\'e}leuc}, \citenamefont {Lienhard},
  \citenamefont {Lahaye},\ and\ \citenamefont {Browaeys}}]{barredo2016atom}%
  \BibitemOpen
  \bibfield  {author} {\bibinfo {author} {\bibfnamefont {D.}~\bibnamefont
  {Barredo}}, \bibinfo {author} {\bibfnamefont {S.}~\bibnamefont
  {De~L{\'e}s{\'e}leuc}}, \bibinfo {author} {\bibfnamefont {V.}~\bibnamefont
  {Lienhard}}, \bibinfo {author} {\bibfnamefont {T.}~\bibnamefont {Lahaye}},\
  and\ \bibinfo {author} {\bibfnamefont {A.}~\bibnamefont {Browaeys}},\
  }\bibfield  {title} {\bibinfo {title} {An atom-by-atom assembler of
  defect-free arbitrary two-dimensional atomic arrays},\ }\href@noop {}
  {\bibfield  {journal} {\bibinfo  {journal} {Science}\ }\textbf {\bibinfo
  {volume} {354}},\ \bibinfo {pages} {1021} (\bibinfo {year}
  {2016})}\BibitemShut {NoStop}%
\bibitem [{\citenamefont {Bernien}\ \emph {et~al.}(2017)\citenamefont
  {Bernien}, \citenamefont {Schwartz}, \citenamefont {Keesling}, \citenamefont
  {Levine}, \citenamefont {Omran}, \citenamefont {Pichler}, \citenamefont
  {Choi}, \citenamefont {Zibrov}, \citenamefont {Endres}, \citenamefont
  {Greiner} \emph {et~al.}}]{bernien2017probing}%
  \BibitemOpen
  \bibfield  {author} {\bibinfo {author} {\bibfnamefont {H.}~\bibnamefont
  {Bernien}}, \bibinfo {author} {\bibfnamefont {S.}~\bibnamefont {Schwartz}},
  \bibinfo {author} {\bibfnamefont {A.}~\bibnamefont {Keesling}}, \bibinfo
  {author} {\bibfnamefont {H.}~\bibnamefont {Levine}}, \bibinfo {author}
  {\bibfnamefont {A.}~\bibnamefont {Omran}}, \bibinfo {author} {\bibfnamefont
  {H.}~\bibnamefont {Pichler}}, \bibinfo {author} {\bibfnamefont
  {S.}~\bibnamefont {Choi}}, \bibinfo {author} {\bibfnamefont {A.~S.}\
  \bibnamefont {Zibrov}}, \bibinfo {author} {\bibfnamefont {M.}~\bibnamefont
  {Endres}}, \bibinfo {author} {\bibfnamefont {M.}~\bibnamefont {Greiner}},
  \emph {et~al.},\ }\bibfield  {title} {\bibinfo {title} {Probing many-body
  dynamics on a 51-atom quantum simulator},\ }\href@noop {} {\bibfield
  {journal} {\bibinfo  {journal} {Nature}\ }\textbf {\bibinfo {volume} {551}},\
  \bibinfo {pages} {579} (\bibinfo {year} {2017})}\BibitemShut {NoStop}%
\bibitem [{\citenamefont {Anderegg}\ \emph {et~al.}(2019)\citenamefont
  {Anderegg}, \citenamefont {Cheuk}, \citenamefont {Bao}, \citenamefont
  {Burchesky}, \citenamefont {Ketterle}, \citenamefont {Ni},\ and\
  \citenamefont {Doyle}}]{anderegg2019optical}%
  \BibitemOpen
  \bibfield  {author} {\bibinfo {author} {\bibfnamefont {L.}~\bibnamefont
  {Anderegg}}, \bibinfo {author} {\bibfnamefont {L.~W.}\ \bibnamefont {Cheuk}},
  \bibinfo {author} {\bibfnamefont {Y.}~\bibnamefont {Bao}}, \bibinfo {author}
  {\bibfnamefont {S.}~\bibnamefont {Burchesky}}, \bibinfo {author}
  {\bibfnamefont {W.}~\bibnamefont {Ketterle}}, \bibinfo {author}
  {\bibfnamefont {K.-K.}\ \bibnamefont {Ni}},\ and\ \bibinfo {author}
  {\bibfnamefont {J.~M.}\ \bibnamefont {Doyle}},\ }\bibfield  {title} {\bibinfo
  {title} {An optical tweezer array of ultracold molecules},\ }\href@noop {}
  {\bibfield  {journal} {\bibinfo  {journal} {Science}\ }\textbf {\bibinfo
  {volume} {365}},\ \bibinfo {pages} {1156} (\bibinfo {year}
  {2019})}\BibitemShut {NoStop}%
\bibitem [{\citenamefont {Zhang}\ \emph {et~al.}(2022)\citenamefont {Zhang},
  \citenamefont {Picard}, \citenamefont {Cairncross}, \citenamefont {Wang},
  \citenamefont {Yu}, \citenamefont {Fang},\ and\ \citenamefont
  {Ni}}]{zhang2022optical}%
  \BibitemOpen
  \bibfield  {author} {\bibinfo {author} {\bibfnamefont {J.~T.}\ \bibnamefont
  {Zhang}}, \bibinfo {author} {\bibfnamefont {L.~R.}\ \bibnamefont {Picard}},
  \bibinfo {author} {\bibfnamefont {W.~B.}\ \bibnamefont {Cairncross}},
  \bibinfo {author} {\bibfnamefont {K.}~\bibnamefont {Wang}}, \bibinfo {author}
  {\bibfnamefont {Y.}~\bibnamefont {Yu}}, \bibinfo {author} {\bibfnamefont
  {F.}~\bibnamefont {Fang}},\ and\ \bibinfo {author} {\bibfnamefont {K.-K.}\
  \bibnamefont {Ni}},\ }\bibfield  {title} {\bibinfo {title} {An optical
  tweezer array of ground-state polar molecules},\ }\href@noop {} {\bibfield
  {journal} {\bibinfo  {journal} {Quantum Science and Technology}\ }\textbf
  {\bibinfo {volume} {7}},\ \bibinfo {pages} {035006} (\bibinfo {year}
  {2022})}\BibitemShut {NoStop}%
\bibitem [{\citenamefont {Park}\ \emph {et~al.}(2017)\citenamefont {Park},
  \citenamefont {Yan}, \citenamefont {Loh}, \citenamefont {Will},\ and\
  \citenamefont {Zwierlein}}]{park2017second}%
  \BibitemOpen
  \bibfield  {author} {\bibinfo {author} {\bibfnamefont {J.~W.}\ \bibnamefont
  {Park}}, \bibinfo {author} {\bibfnamefont {Z.~Z.}\ \bibnamefont {Yan}},
  \bibinfo {author} {\bibfnamefont {H.}~\bibnamefont {Loh}}, \bibinfo {author}
  {\bibfnamefont {S.~A.}\ \bibnamefont {Will}},\ and\ \bibinfo {author}
  {\bibfnamefont {M.~W.}\ \bibnamefont {Zwierlein}},\ }\bibfield  {title}
  {\bibinfo {title} {Second-scale nuclear spin coherence time of ultracold
  $^{23}\text{Na} ^{40}\text{K}$ molecules},\ }\href@noop {} {\bibfield
  {journal} {\bibinfo  {journal} {Science}\ }\textbf {\bibinfo {volume}
  {357}},\ \bibinfo {pages} {372} (\bibinfo {year} {2017})}\BibitemShut
  {NoStop}%
\bibitem [{\citenamefont {Gregory}\ \emph {et~al.}(2021)\citenamefont
  {Gregory}, \citenamefont {Blackmore}, \citenamefont {Bromley}, \citenamefont
  {Hutson},\ and\ \citenamefont {Cornish}}]{gregory2021robust}%
  \BibitemOpen
  \bibfield  {author} {\bibinfo {author} {\bibfnamefont {P.~D.}\ \bibnamefont
  {Gregory}}, \bibinfo {author} {\bibfnamefont {J.~A.}\ \bibnamefont
  {Blackmore}}, \bibinfo {author} {\bibfnamefont {S.~L.}\ \bibnamefont
  {Bromley}}, \bibinfo {author} {\bibfnamefont {J.~M.}\ \bibnamefont
  {Hutson}},\ and\ \bibinfo {author} {\bibfnamefont {S.~L.}\ \bibnamefont
  {Cornish}},\ }\bibfield  {title} {\bibinfo {title} {Robust storage qubits in
  ultracold polar molecules},\ }\href@noop {} {\bibfield  {journal} {\bibinfo
  {journal} {Nature Physics}\ }\textbf {\bibinfo {volume} {17}},\ \bibinfo
  {pages} {1149} (\bibinfo {year} {2021})}\BibitemShut {NoStop}%
\bibitem [{\citenamefont {Gorshkov}\ \emph {et~al.}(2011)\citenamefont
  {Gorshkov}, \citenamefont {Manmana}, \citenamefont {Chen}, \citenamefont
  {Ye}, \citenamefont {Demler}, \citenamefont {Lukin},\ and\ \citenamefont
  {Rey}}]{gorshkov2011tunable}%
  \BibitemOpen
  \bibfield  {author} {\bibinfo {author} {\bibfnamefont {A.~V.}\ \bibnamefont
  {Gorshkov}}, \bibinfo {author} {\bibfnamefont {S.~R.}\ \bibnamefont
  {Manmana}}, \bibinfo {author} {\bibfnamefont {G.}~\bibnamefont {Chen}},
  \bibinfo {author} {\bibfnamefont {J.}~\bibnamefont {Ye}}, \bibinfo {author}
  {\bibfnamefont {E.}~\bibnamefont {Demler}}, \bibinfo {author} {\bibfnamefont
  {M.~D.}\ \bibnamefont {Lukin}},\ and\ \bibinfo {author} {\bibfnamefont
  {A.~M.}\ \bibnamefont {Rey}},\ }\bibfield  {title} {\bibinfo {title} {Tunable
  superfluidity and quantum magnetism with ultracold polar molecules},\
  }\href@noop {} {\bibfield  {journal} {\bibinfo  {journal} {Physical review
  letters}\ }\textbf {\bibinfo {volume} {107}},\ \bibinfo {pages} {115301}
  (\bibinfo {year} {2011})}\BibitemShut {NoStop}%
\bibitem [{\citenamefont {Tobias}\ \emph {et~al.}(2022)\citenamefont {Tobias},
  \citenamefont {Matsuda}, \citenamefont {Li}, \citenamefont {Miller},
  \citenamefont {Carroll}, \citenamefont {Bilitewski}, \citenamefont {Rey},\
  and\ \citenamefont {Ye}}]{tobias2022reactions}%
  \BibitemOpen
  \bibfield  {author} {\bibinfo {author} {\bibfnamefont {W.~G.}\ \bibnamefont
  {Tobias}}, \bibinfo {author} {\bibfnamefont {K.}~\bibnamefont {Matsuda}},
  \bibinfo {author} {\bibfnamefont {J.-R.}\ \bibnamefont {Li}}, \bibinfo
  {author} {\bibfnamefont {C.}~\bibnamefont {Miller}}, \bibinfo {author}
  {\bibfnamefont {A.~N.}\ \bibnamefont {Carroll}}, \bibinfo {author}
  {\bibfnamefont {T.}~\bibnamefont {Bilitewski}}, \bibinfo {author}
  {\bibfnamefont {A.~M.}\ \bibnamefont {Rey}},\ and\ \bibinfo {author}
  {\bibfnamefont {J.}~\bibnamefont {Ye}},\ }\bibfield  {title} {\bibinfo
  {title} {Reactions between layer-resolved molecules mediated by dipolar spin
  exchange},\ }\href@noop {} {\bibfield  {journal} {\bibinfo  {journal}
  {Science}\ }\textbf {\bibinfo {volume} {375}},\ \bibinfo {pages} {1299}
  (\bibinfo {year} {2022})}\BibitemShut {NoStop}%
\bibitem [{\citenamefont {Holland}\ \emph {et~al.}(2022)\citenamefont
  {Holland}, \citenamefont {Lu},\ and\ \citenamefont
  {Cheuk}}]{holland2022demand}%
  \BibitemOpen
  \bibfield  {author} {\bibinfo {author} {\bibfnamefont {C.~M.}\ \bibnamefont
  {Holland}}, \bibinfo {author} {\bibfnamefont {Y.}~\bibnamefont {Lu}},\ and\
  \bibinfo {author} {\bibfnamefont {L.~W.}\ \bibnamefont {Cheuk}},\ }\bibfield
  {title} {\bibinfo {title} {On-demand entanglement of molecules in a
  reconfigurable optical tweezer array},\ }\href@noop {} {\bibfield  {journal}
  {\bibinfo  {journal} {arXiv preprint arXiv:2210.06309}\ } (\bibinfo {year}
  {2022})}\BibitemShut {NoStop}%
\bibitem [{\citenamefont {Turchette}\ \emph {et~al.}(1998)\citenamefont
  {Turchette}, \citenamefont {Wood}, \citenamefont {King}, \citenamefont
  {Myatt}, \citenamefont {Leibfried}, \citenamefont {Itano}, \citenamefont
  {Monroe},\ and\ \citenamefont {Wineland}}]{turchette1998deterministic}%
  \BibitemOpen
  \bibfield  {author} {\bibinfo {author} {\bibfnamefont {Q.}~\bibnamefont
  {Turchette}}, \bibinfo {author} {\bibfnamefont {C.}~\bibnamefont {Wood}},
  \bibinfo {author} {\bibfnamefont {B.}~\bibnamefont {King}}, \bibinfo {author}
  {\bibfnamefont {C.}~\bibnamefont {Myatt}}, \bibinfo {author} {\bibfnamefont
  {D.}~\bibnamefont {Leibfried}}, \bibinfo {author} {\bibfnamefont
  {W.}~\bibnamefont {Itano}}, \bibinfo {author} {\bibfnamefont
  {C.}~\bibnamefont {Monroe}},\ and\ \bibinfo {author} {\bibfnamefont
  {D.}~\bibnamefont {Wineland}},\ }\bibfield  {title} {\bibinfo {title}
  {Deterministic entanglement of two trapped ions},\ }\href@noop {} {\bibfield
  {journal} {\bibinfo  {journal} {Physical Review Letters}\ }\textbf {\bibinfo
  {volume} {81}},\ \bibinfo {pages} {3631} (\bibinfo {year}
  {1998})}\BibitemShut {NoStop}%
\bibitem [{\citenamefont {Caldwell}\ and\ \citenamefont
  {Tarbutt}(2020)}]{caldwell2020sideband}%
  \BibitemOpen
  \bibfield  {author} {\bibinfo {author} {\bibfnamefont {L.}~\bibnamefont
  {Caldwell}}\ and\ \bibinfo {author} {\bibfnamefont {M.}~\bibnamefont
  {Tarbutt}},\ }\bibfield  {title} {\bibinfo {title} {Sideband cooling of
  molecules in optical traps},\ }\href@noop {} {\bibfield  {journal} {\bibinfo
  {journal} {Physical Review Research}\ }\textbf {\bibinfo {volume} {2}},\
  \bibinfo {pages} {013251} (\bibinfo {year} {2020})}\BibitemShut {NoStop}%
\bibitem [{\citenamefont {Hutzler}\ \emph {et~al.}(2012)\citenamefont
  {Hutzler}, \citenamefont {Lu},\ and\ \citenamefont
  {Doyle}}]{hutzler2012buffer}%
  \BibitemOpen
  \bibfield  {author} {\bibinfo {author} {\bibfnamefont {N.~R.}\ \bibnamefont
  {Hutzler}}, \bibinfo {author} {\bibfnamefont {H.-I.}\ \bibnamefont {Lu}},\
  and\ \bibinfo {author} {\bibfnamefont {J.~M.}\ \bibnamefont {Doyle}},\
  }\bibfield  {title} {\bibinfo {title} {The buffer gas beam: An intense, cold,
  and slow source for atoms and molecules},\ }\href@noop {} {\bibfield
  {journal} {\bibinfo  {journal} {Chemical Reviews}\ }\textbf {\bibinfo
  {volume} {112}},\ \bibinfo {pages} {4803} (\bibinfo {year}
  {2012})}\BibitemShut {NoStop}%
\bibitem [{\citenamefont {Cheuk}\ \emph {et~al.}(2018)\citenamefont {Cheuk},
  \citenamefont {Anderegg}, \citenamefont {Augenbraun}, \citenamefont {Bao},
  \citenamefont {Burchesky}, \citenamefont {Ketterle},\ and\ \citenamefont
  {Doyle}}]{cheuk2018lambda}%
  \BibitemOpen
  \bibfield  {author} {\bibinfo {author} {\bibfnamefont {L.~W.}\ \bibnamefont
  {Cheuk}}, \bibinfo {author} {\bibfnamefont {L.}~\bibnamefont {Anderegg}},
  \bibinfo {author} {\bibfnamefont {B.~L.}\ \bibnamefont {Augenbraun}},
  \bibinfo {author} {\bibfnamefont {Y.}~\bibnamefont {Bao}}, \bibinfo {author}
  {\bibfnamefont {S.}~\bibnamefont {Burchesky}}, \bibinfo {author}
  {\bibfnamefont {W.}~\bibnamefont {Ketterle}},\ and\ \bibinfo {author}
  {\bibfnamefont {J.~M.}\ \bibnamefont {Doyle}},\ }\bibfield  {title} {\bibinfo
  {title} {{$\Lambda$}-enhanced imaging of molecules in an optical trap},\
  }\href@noop {} {\bibfield  {journal} {\bibinfo  {journal} {Physical Review
  Letters}\ }\textbf {\bibinfo {volume} {121}},\ \bibinfo {pages} {083201}
  (\bibinfo {year} {2018})}\BibitemShut {NoStop}%
\bibitem [{\citenamefont {Anderegg}\ \emph {et~al.}(2018)\citenamefont
  {Anderegg}, \citenamefont {Augenbraun}, \citenamefont {Bao}, \citenamefont
  {Burchesky}, \citenamefont {Cheuk}, \citenamefont {Ketterle},\ and\
  \citenamefont {Doyle}}]{anderegg2018laser}%
  \BibitemOpen
  \bibfield  {author} {\bibinfo {author} {\bibfnamefont {L.}~\bibnamefont
  {Anderegg}}, \bibinfo {author} {\bibfnamefont {B.~L.}\ \bibnamefont
  {Augenbraun}}, \bibinfo {author} {\bibfnamefont {Y.}~\bibnamefont {Bao}},
  \bibinfo {author} {\bibfnamefont {S.}~\bibnamefont {Burchesky}}, \bibinfo
  {author} {\bibfnamefont {L.~W.}\ \bibnamefont {Cheuk}}, \bibinfo {author}
  {\bibfnamefont {W.}~\bibnamefont {Ketterle}},\ and\ \bibinfo {author}
  {\bibfnamefont {J.~M.}\ \bibnamefont {Doyle}},\ }\bibfield  {title} {\bibinfo
  {title} {Laser cooling of optically trapped molecules},\ }\href@noop {}
  {\bibfield  {journal} {\bibinfo  {journal} {Nature Physics}\ }\textbf
  {\bibinfo {volume} {14}},\ \bibinfo {pages} {890} (\bibinfo {year}
  {2018})}\BibitemShut {NoStop}%
\bibitem [{\citenamefont {Bao}\ \emph {et~al.}(2022)\citenamefont {Bao},
  \citenamefont {Scarlett}, \citenamefont {Anderegg}, \citenamefont
  {Burchesky}, \citenamefont {Gonzalez-Acevedo}, \citenamefont {Chae},
  \citenamefont {Ketterle}, \citenamefont {Ni},\ and\ \citenamefont
  {Doyle}}]{bao2022fast}%
  \BibitemOpen
  \bibfield  {author} {\bibinfo {author} {\bibfnamefont {Y.}~\bibnamefont
  {Bao}}, \bibinfo {author} {\bibfnamefont {S.~Y.}\ \bibnamefont {Scarlett}},
  \bibinfo {author} {\bibfnamefont {L.}~\bibnamefont {Anderegg}}, \bibinfo
  {author} {\bibfnamefont {S.}~\bibnamefont {Burchesky}}, \bibinfo {author}
  {\bibfnamefont {D.}~\bibnamefont {Gonzalez-Acevedo}}, \bibinfo {author}
  {\bibfnamefont {E.}~\bibnamefont {Chae}}, \bibinfo {author} {\bibfnamefont
  {W.}~\bibnamefont {Ketterle}}, \bibinfo {author} {\bibfnamefont {K.-K.}\
  \bibnamefont {Ni}},\ and\ \bibinfo {author} {\bibfnamefont {J.~M.}\
  \bibnamefont {Doyle}},\ }\bibfield  {title} {\bibinfo {title} {Fast optical
  transport of ultracold molecules over long distances},\ }\href@noop {}
  {\bibfield  {journal} {\bibinfo  {journal} {New Journal of Physics}\ }\textbf
  {\bibinfo {volume} {24}},\ \bibinfo {pages} {093028} (\bibinfo {year}
  {2022})}\BibitemShut {NoStop}%
\bibitem [{\citenamefont {Neyenhuis}\ \emph {et~al.}(2012)\citenamefont
  {Neyenhuis}, \citenamefont {Yan}, \citenamefont {Moses}, \citenamefont
  {Covey}, \citenamefont {Chotia}, \citenamefont {Petrov}, \citenamefont
  {Kotochigova}, \citenamefont {Ye},\ and\ \citenamefont
  {Jin}}]{neyenhuis2012anisotropic}%
  \BibitemOpen
  \bibfield  {author} {\bibinfo {author} {\bibfnamefont {B.}~\bibnamefont
  {Neyenhuis}}, \bibinfo {author} {\bibfnamefont {B.}~\bibnamefont {Yan}},
  \bibinfo {author} {\bibfnamefont {S.}~\bibnamefont {Moses}}, \bibinfo
  {author} {\bibfnamefont {J.}~\bibnamefont {Covey}}, \bibinfo {author}
  {\bibfnamefont {A.}~\bibnamefont {Chotia}}, \bibinfo {author} {\bibfnamefont
  {A.}~\bibnamefont {Petrov}}, \bibinfo {author} {\bibfnamefont
  {S.}~\bibnamefont {Kotochigova}}, \bibinfo {author} {\bibfnamefont
  {J.}~\bibnamefont {Ye}},\ and\ \bibinfo {author} {\bibfnamefont
  {D.}~\bibnamefont {Jin}},\ }\bibfield  {title} {\bibinfo {title} {Anisotropic
  polarizability of ultracold polar $^{40}\text{K} ^{87}\text{Rb}$ molecules},\
  }\href@noop {} {\bibfield  {journal} {\bibinfo  {journal} {Physical Review
  Letters}\ }\textbf {\bibinfo {volume} {109}},\ \bibinfo {pages} {230403}
  (\bibinfo {year} {2012})}\BibitemShut {NoStop}%
\bibitem [{\citenamefont {Chae}(2021)}]{chae2021entanglement}%
  \BibitemOpen
  \bibfield  {author} {\bibinfo {author} {\bibfnamefont {E.}~\bibnamefont
  {Chae}},\ }\bibfield  {title} {\bibinfo {title} {Entanglement via rotational
  blockade of mgf molecules in a magic potential},\ }\href@noop {} {\bibfield
  {journal} {\bibinfo  {journal} {Physical Chemistry Chemical Physics}\
  }\textbf {\bibinfo {volume} {23}},\ \bibinfo {pages} {1215} (\bibinfo {year}
  {2021})}\BibitemShut {NoStop}%
\bibitem [{\citenamefont {Lin}\ \emph {et~al.}(2021)\citenamefont {Lin},
  \citenamefont {He}, \citenamefont {Ye},\ and\ \citenamefont
  {Wang}}]{lin2021anisotropic}%
  \BibitemOpen
  \bibfield  {author} {\bibinfo {author} {\bibfnamefont {J.}~\bibnamefont
  {Lin}}, \bibinfo {author} {\bibfnamefont {J.}~\bibnamefont {He}}, \bibinfo
  {author} {\bibfnamefont {X.}~\bibnamefont {Ye}},\ and\ \bibinfo {author}
  {\bibfnamefont {D.}~\bibnamefont {Wang}},\ }\bibfield  {title} {\bibinfo
  {title} {Anisotropic polarizability of ultracold ground-state $^{23}\text{Na}
  ^{87}\text{Rb}$ molecules},\ }\href@noop {} {\bibfield  {journal} {\bibinfo
  {journal} {Physical Review A}\ }\textbf {\bibinfo {volume} {103}},\ \bibinfo
  {pages} {023332} (\bibinfo {year} {2021})}\BibitemShut {NoStop}%
\bibitem [{\citenamefont {Gullion}\ \emph {et~al.}(1990)\citenamefont
  {Gullion}, \citenamefont {Baker},\ and\ \citenamefont
  {Conradi}}]{gullion1990new}%
  \BibitemOpen
  \bibfield  {author} {\bibinfo {author} {\bibfnamefont {T.}~\bibnamefont
  {Gullion}}, \bibinfo {author} {\bibfnamefont {D.~B.}\ \bibnamefont {Baker}},\
  and\ \bibinfo {author} {\bibfnamefont {M.~S.}\ \bibnamefont {Conradi}},\
  }\bibfield  {title} {\bibinfo {title} {New, compensated carr-ourcell
  sequences},\ }\href@noop {} {\bibfield  {journal} {\bibinfo  {journal}
  {Journal of Magnetic Resonance (1969)}\ }\textbf {\bibinfo {volume} {89}},\
  \bibinfo {pages} {479} (\bibinfo {year} {1990})}\BibitemShut {NoStop}%
\bibitem [{\citenamefont {Du}\ \emph {et~al.}(2009)\citenamefont {Du},
  \citenamefont {Rong}, \citenamefont {Zhao}, \citenamefont {Wang},
  \citenamefont {Yang},\ and\ \citenamefont {Liu}}]{du2009preserving}%
  \BibitemOpen
  \bibfield  {author} {\bibinfo {author} {\bibfnamefont {J.}~\bibnamefont
  {Du}}, \bibinfo {author} {\bibfnamefont {X.}~\bibnamefont {Rong}}, \bibinfo
  {author} {\bibfnamefont {N.}~\bibnamefont {Zhao}}, \bibinfo {author}
  {\bibfnamefont {Y.}~\bibnamefont {Wang}}, \bibinfo {author} {\bibfnamefont
  {J.}~\bibnamefont {Yang}},\ and\ \bibinfo {author} {\bibfnamefont
  {R.}~\bibnamefont {Liu}},\ }\bibfield  {title} {\bibinfo {title} {Preserving
  electron spin coherence in solids by optimal dynamical decoupling},\
  }\href@noop {} {\bibfield  {journal} {\bibinfo  {journal} {Nature}\ }\textbf
  {\bibinfo {volume} {461}},\ \bibinfo {pages} {1265} (\bibinfo {year}
  {2009})}\BibitemShut {NoStop}%
\bibitem [{\citenamefont {Choi}\ \emph {et~al.}(2020)\citenamefont {Choi},
  \citenamefont {Zhou}, \citenamefont {Knowles}, \citenamefont {Landig},
  \citenamefont {Choi},\ and\ \citenamefont {Lukin}}]{choi2020robust}%
  \BibitemOpen
  \bibfield  {author} {\bibinfo {author} {\bibfnamefont {J.}~\bibnamefont
  {Choi}}, \bibinfo {author} {\bibfnamefont {H.}~\bibnamefont {Zhou}}, \bibinfo
  {author} {\bibfnamefont {H.~S.}\ \bibnamefont {Knowles}}, \bibinfo {author}
  {\bibfnamefont {R.}~\bibnamefont {Landig}}, \bibinfo {author} {\bibfnamefont
  {S.}~\bibnamefont {Choi}},\ and\ \bibinfo {author} {\bibfnamefont {M.~D.}\
  \bibnamefont {Lukin}},\ }\bibfield  {title} {\bibinfo {title} {Robust dynamic
  hamiltonian engineering of many-body spin systems},\ }\href@noop {}
  {\bibfield  {journal} {\bibinfo  {journal} {Physical Review X}\ }\textbf
  {\bibinfo {volume} {10}},\ \bibinfo {pages} {031002} (\bibinfo {year}
  {2020})}\BibitemShut {NoStop}%
\bibitem [{\citenamefont {Sackett}\ \emph {et~al.}(2000)\citenamefont
  {Sackett}, \citenamefont {Kielpinski}, \citenamefont {King}, \citenamefont
  {Langer}, \citenamefont {Meyer}, \citenamefont {Myatt}, \citenamefont {Rowe},
  \citenamefont {Turchette}, \citenamefont {Itano}, \citenamefont {Wineland}
  \emph {et~al.}}]{sackett2000experimental}%
  \BibitemOpen
  \bibfield  {author} {\bibinfo {author} {\bibfnamefont {C.~A.}\ \bibnamefont
  {Sackett}}, \bibinfo {author} {\bibfnamefont {D.}~\bibnamefont {Kielpinski}},
  \bibinfo {author} {\bibfnamefont {B.~E.}\ \bibnamefont {King}}, \bibinfo
  {author} {\bibfnamefont {C.}~\bibnamefont {Langer}}, \bibinfo {author}
  {\bibfnamefont {V.}~\bibnamefont {Meyer}}, \bibinfo {author} {\bibfnamefont
  {C.~J.}\ \bibnamefont {Myatt}}, \bibinfo {author} {\bibfnamefont
  {M.}~\bibnamefont {Rowe}}, \bibinfo {author} {\bibfnamefont {Q.}~\bibnamefont
  {Turchette}}, \bibinfo {author} {\bibfnamefont {W.~M.}\ \bibnamefont
  {Itano}}, \bibinfo {author} {\bibfnamefont {D.~J.}\ \bibnamefont {Wineland}},
  \emph {et~al.},\ }\bibfield  {title} {\bibinfo {title} {Experimental
  entanglement of four particles},\ }\href@noop {} {\bibfield  {journal}
  {\bibinfo  {journal} {Nature}\ }\textbf {\bibinfo {volume} {404}},\ \bibinfo
  {pages} {256} (\bibinfo {year} {2000})}\BibitemShut {NoStop}%
\bibitem [{\citenamefont {Pezze}\ \emph {et~al.}(2018)\citenamefont {Pezze},
  \citenamefont {Smerzi}, \citenamefont {Oberthaler}, \citenamefont {Schmied},\
  and\ \citenamefont {Treutlein}}]{pezze2018quantum}%
  \BibitemOpen
  \bibfield  {author} {\bibinfo {author} {\bibfnamefont {L.}~\bibnamefont
  {Pezze}}, \bibinfo {author} {\bibfnamefont {A.}~\bibnamefont {Smerzi}},
  \bibinfo {author} {\bibfnamefont {M.~K.}\ \bibnamefont {Oberthaler}},
  \bibinfo {author} {\bibfnamefont {R.}~\bibnamefont {Schmied}},\ and\ \bibinfo
  {author} {\bibfnamefont {P.}~\bibnamefont {Treutlein}},\ }\bibfield  {title}
  {\bibinfo {title} {Quantum metrology with nonclassical states of atomic
  ensembles},\ }\href@noop {} {\bibfield  {journal} {\bibinfo  {journal}
  {Reviews of Modern Physics}\ }\textbf {\bibinfo {volume} {90}},\ \bibinfo
  {pages} {035005} (\bibinfo {year} {2018})}\BibitemShut {NoStop}%
\bibitem [{\citenamefont {Yu}\ \emph {et~al.}(2019)\citenamefont {Yu},
  \citenamefont {Cheuk}, \citenamefont {Kozyryev},\ and\ \citenamefont
  {Doyle}}]{yu2019scalable}%
  \BibitemOpen
  \bibfield  {author} {\bibinfo {author} {\bibfnamefont {P.}~\bibnamefont
  {Yu}}, \bibinfo {author} {\bibfnamefont {L.~W.}\ \bibnamefont {Cheuk}},
  \bibinfo {author} {\bibfnamefont {I.}~\bibnamefont {Kozyryev}},\ and\
  \bibinfo {author} {\bibfnamefont {J.~M.}\ \bibnamefont {Doyle}},\ }\bibfield
  {title} {\bibinfo {title} {A scalable quantum computing platform using
  symmetric-top molecules},\ }\href@noop {} {\bibfield  {journal} {\bibinfo
  {journal} {New Journal of Physics}\ }\textbf {\bibinfo {volume} {21}},\
  \bibinfo {pages} {093049} (\bibinfo {year} {2019})}\BibitemShut {NoStop}%
\bibitem [{\citenamefont {Hallas}\ \emph {et~al.}(2022)\citenamefont {Hallas},
  \citenamefont {Vilas}, \citenamefont {Anderegg}, \citenamefont {Robichaud},
  \citenamefont {Winnicki}, \citenamefont {Zhang}, \citenamefont {Cheng},\ and\
  \citenamefont {Doyle}}]{hallas2022optical}%
  \BibitemOpen
  \bibfield  {author} {\bibinfo {author} {\bibfnamefont {C.}~\bibnamefont
  {Hallas}}, \bibinfo {author} {\bibfnamefont {N.~B.}\ \bibnamefont {Vilas}},
  \bibinfo {author} {\bibfnamefont {L.}~\bibnamefont {Anderegg}}, \bibinfo
  {author} {\bibfnamefont {P.}~\bibnamefont {Robichaud}}, \bibinfo {author}
  {\bibfnamefont {A.}~\bibnamefont {Winnicki}}, \bibinfo {author}
  {\bibfnamefont {C.}~\bibnamefont {Zhang}}, \bibinfo {author} {\bibfnamefont
  {L.}~\bibnamefont {Cheng}},\ and\ \bibinfo {author} {\bibfnamefont {J.~M.}\
  \bibnamefont {Doyle}},\ }\bibfield  {title} {\bibinfo {title} {Optical
  trapping of a polyatomic molecule in an $\ell$-type parity doublet state},\
  }\href@noop {} {\bibfield  {journal} {\bibinfo  {journal} {arXiv preprint
  arXiv:2208.13762}\ } (\bibinfo {year} {2022})}\BibitemShut {NoStop}%
\end{thebibliography}%

\end{document}